\documentclass[twocolumn,aps,prb,superscriptaddress,notitlepage,longbibliography]{revtex4}
\usepackage[colorlinks=true,urlcolor=blue,citecolor=blue,linkcolor=blue]{hyperref}
\usepackage{graphicx}  
\usepackage{dcolumn}   
\usepackage{bm}        
\usepackage{amssymb}
\usepackage{amsfonts}
\usepackage{verbatim}
\usepackage{epstopdf}

\usepackage{amsmath}
\usepackage{braket}
\usepackage{hyperref}
\DeclareMathOperator{\Tr}{Tr}

\begin{document}

\title{ Disorder and quantum transport of helical quantum Hall phase in graphene}
\author{Yue-Ran Ding}
\affiliation{Institute for Advanced Study and School of Physical Science and Technology, Soochow University, Suzhou 215006, China}
\author{Dong-Hui Xu}\thanks{donghuixu@cqu.edu.cn}
\affiliation{Department of Physics, Chongqing University, Chongqing 400044, China}
\affiliation{Chongqing Key Laboratory for Strongly Coupled Physics, Chongqing University, Chongqing 400044, China}
\author{Chui-Zhen Chen}\thanks{czchen@suda.edu.cn}
\affiliation{Institute for Advanced Study and School of Physical Science and Technology, Soochow University, Suzhou 215006, China}

\begin{abstract}
Recently, an exotic quantum Hall ferromagnet with spin-filtered helical edge modes was observed in monolayer graphene on a high-dielectric constant substrate at moderate magnetic fields, withstanding temperatures of up to 110 Kelvin~[L. Veyrat et al., Science 367, 781 (2020)]. However, the characteristic quantized longitudinal resistance mediated by these edge modes departs from quantization with decreasing temperature. 
 In this work, we investigate the transport properties of helical edge modes in a graphene nanoribbon under a perpendicular magnetic field using the Landauer-Buttiker transport formalism. We find that the departure of quantization of longitudinal conductance is due to the helical-edge gap opened by the Rashba spin-orbital coupling. The quantization can be restored by weak nonmagnetic Anderson disorder at low temperature, increasing the localization length, or by raising temperature at weak disorder, through thermal broadening. 
 The resulted conductance is very close to the quantized value $2e^2/h$, which is in qualitatively consistent with the experimental results.
Furthermore, we suggest that the helical quantum Hall phase in graphene could be a promising platform for creating Majorana zero modes by introducing superconductivity.

\end{abstract}

\maketitle

\section{Introduction}
The quantum spin Hall effect (QSHE) is an intriguing two-dimensional (2D) topological state of matter characterized by a pair of gapless helical edge states within the bulk gap.
These helical edge states, when accompanied by time-reversal symmetry, exhibit remarkable immunity to backscattering, resulting in a distinct quantized edge conductance.
Initially proposed by Kane and Mele, the QSHE was first identified in monolayer graphene, where the presence of intrinsic spin-orbital coupling (SOC) led to the opening of a band gap and the emergence of edge states~\cite{KaneMelePRL20051,KaneMelePRL20052}. However, due to the weak SOC in graphene, observing the QSHE in this material proved challenging~\cite{HKMinPRB2006,YYgPRB2007}.
Subsequently, the QSHE was predicted to exist in other systems such as the HgTe/CdTe quantum well and InAs/GaSb quantum well, as well as in the 1T$^\prime$-WTe2 monolayer ~\cite{ShengLPRL2005,BernevigPRL2006,NurakamiPRL2006,FuLiangPRL2007,LiuChaoxingPRL2008,HsiehDN2008}. Nevertheless, the direct experimental observation of the hallmark quantized edge conductance plateau in the QSHE has thus far been limited to a few specific materials at very low temperatures (a few Kelvin) \cite{KonigS2007,KnezPRL2011,FeiZNP2017,HatsudaSA2018,WSFS2018}.

On the other side, graphene, a monolayer honeycomb lattice of carbon atoms, has been attracted enormous interest in the scientific community over the past two decades \cite{NovoselovS2004,NovoselovN2005,NovoselovNP2006,ZhangYBN2005,BeenakkerRMP2008,NetoRMP2009}. As a topological semimetal, graphene exhibits a unique electronic band structure where the conduction and valence bands linearly intersect at Dirac points located at the K and K$^{\prime}$ valleys. Quasiparticles in graphene obey a 2D analogue of the Dirac equation, lending to their relativistic-like transport behavior.
One of the most striking examples of this behavior is the unconventional integer quantum Hall effect (IQHE) observed in graphene under a perpendicular magnetic field. This IQHE is caused by the quantum anomaly of the $n = 0$ Landau level. The Zeeman spin splitting, which can be enhanced by exchange interaction, removes the spin degeneracy of the Landau level and can induce a topological inversion between the electron-type and hole-type $n = 0$ Landau level \cite{BLPRB2006,ADAPRL2006,FHAPRL2006}. Consequently, a pair of counterpropagating, spin-filtered helical edge channels emerges between the two inverted Landau levels, similar to those observed in the QSHE. Such a spin-polarized ferromagnetic phase with zero Chern number is termed a quantum Hall topological insulator \cite{KharitonovPRB2016} or helical quantum Hall phase \cite{VeyratScience2020}, and it
is protected by $CT$ invariance (where $C$ is the charge conjugation operation and $T$ is the time-reversal operation) \cite{QFSunPRL2010}. 

Remarkably, a recent experiment observed a quantized longitudinal resistance in the helical quantum Hall phase of monolayer graphene under moderate perpendicular magnetic fields, reaching temperatures of up to 110 Kelvin~\cite{VeyratScience2020}. However, as the temperature decreases, the quantized two-terminal resistance of graphene increases and deviates from its originally quantized value. This behavior contradicts the transport properties of the QSHE, where the quantized transport is lost as the temperature increases.
To address this discrepancy, it is crucial to consider the Rashba spin-orbit coupling (RSOC) arising from the structural inversion asymmetry induced by the substrate, which has been overlooked in previous studies~\cite{ADAPRL2006,FHAPRL2006}. In practice, the RSOC can be significantly enhanced, as experimentally demonstrated, reaching values on the order of 10 meV~\cite{GMPRB2009}. In the case of graphene deposited on a Ni substrate, the RSOC can even reach values as high as 200 meV~\cite{DYSPRL2008}. Therefore, conducting a systematic investigation of the strong RSOC effect on the transport of helical quantum Hall phase in graphene is of utmost importance.

\begin{figure*}[tbh]
	\centering
	\includegraphics[width=6.8in]{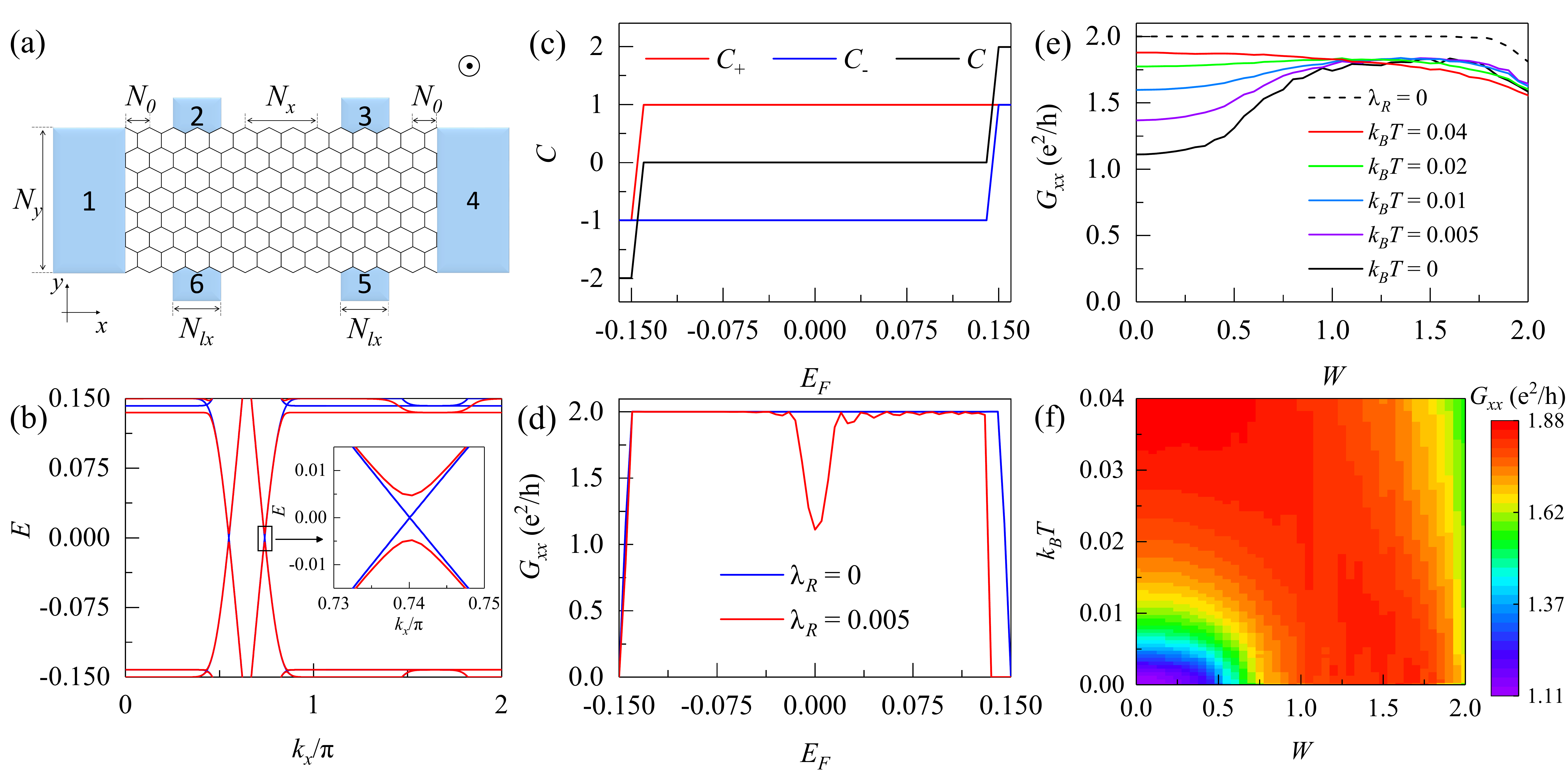}
	\caption{(Color online). (a) Schematic diagram of a six-terminal device. (b) The energy band structure under open boundary condition (OBC) along the $y$-direction, in the presence of a perpendicular magnetic field.  (c) The Chern number C and spin Chern number $C_{\pm}$ versus the chemical potential $E_F$ under PBC with $\phi=\pi/40$ and $\lambda_R=0.005$. (d) The longitudinal conductance $G_{xx}$ as a function of $E_F$. In (b) and (d), RSOC $\lambda_R=0$ and $\lambda_R=0.005$ for blue and red lines, respectively. (e) Dependence of $G_{xx}$ on disorder strength $W$ for various temperatures $k_BT$. $\lambda_R=0$ for dashed black line and $\lambda_R=0.005$ for others. (f) $G_{xx}$ for various $k_BT$ and $W$ at $\lambda_R=0.005$. $G_{xx}$ are averaged over $200$ random configurations. Other parameters are chosen as $N_y=160$, $N_0=100$, $N_{lx}=40$, $N_x=100$, $t=1$, $\phi=0.05$ and $M_z=0.15$.
		\label{fig1} }
\end{figure*}
In this paper, we numerically study the transport properties of the helical quantum Hall phase in a graphene nanoribbon with RSOC subjected to a perpendicular magnetic field.
Specifically, we consider a six-terminal device configuration~\cite{HuaJiangPRL2009,RRDuPRL2015} [see Fig.~\ref{fig1}a] to determine the longitudinal conductance with the aid of the Landauer-Buttiker formalism. The introduction of RSOC results in the opening of an energy gap in the helical edge states, leading to a suppression of the edge conductance and hindering the quantization process. We found that the quantization of edge conductance could be restored in two ways: (i) the application of weak non-magnetic disorder in the low-temperature regime or (ii) an increase in temperature under the influence of weak disorder. To gain insight into the impact of disorder on the edge transport, we employ the transfer matrix method to calculate the localization length. Our calculations reveal that weak non-magnetic disorder increases the localization length, thereby enhancing the conductance. Furthermore, we propose that when in proximity to an $s$-wave superconductor, the helical edge modes in the helical quantum Hall phase of the graphene can naturally host Majorana zero modes.


\section{ Model Hamiltonian}
For a zigzag edged graphene nanoribbon with the RSOC in perpendicular magnetic field,
the tight-binding Hamiltonian can be written as
\begin{eqnarray}
H\!\!=-\sum_{\langle ij\rangle,\sigma}te^{i\phi_{ij}}c_{i}^{\dagger}c_{j}+\sum_ic_i^{\dagger}
M_z\sigma_z c_i\nonumber
\\-\sum_{\langle ij\rangle}i\lambda_Re^{i\phi_{ij}}c_i^{\dagger}(\boldsymbol{\sigma}\times\boldsymbol{\hat{r}}_{ij})_zc_j,
\end{eqnarray}
where $c_i^{\dagger}$ and $c_i$ are the creation and annihilation operators for electrons with two spin components at discrete site $i$ on honeycomb lattice. The Pauli vector $\boldsymbol{\sigma}$ describes the electron's spin, $t$ is the nearest neighbor hopping integral, and $M_z$ denotes the Zeeman splitting. In the presence of a perpendicular magnetic field $B_{\bot}$, a phase factor $\phi_{ij}$ is introduced to the hopping integral, where $\phi_{ij}=\int_i^j{\boldsymbol{A}}\cdot\!d{\boldsymbol{l}}/\Phi_0$ with the vector potential ${\boldsymbol{A}}=(yB_{\bot},0,0)$ and $\Phi_0=\hbar/e$. The magnetic flux in each honeycomb lattice is denoted by $\phi$, where $\phi=(3\sqrt3/4)a^2B_{\bot}/\Phi_0$ \cite{QFSunPRL2010}.
The third term in the Hamiltonian describes the RSOC with a strength $\lambda_R$. Here $\boldsymbol{\hat{r}}_{ij}$ represents the unit vector along the bond that connects the nearest neighbor sites $j$ and $i$ \cite{KaneMelePRL20051}.
In the subsequent calculations, we adopt $t=1$ as the energy unit.

In the absence of the Zeeman spin splitting, the Landau levels in graphene have spin degeneracy. When a Zeeman magnetic field $M_z$ is applied, the spin degeneracy is lifted and the zeroth Landau levels of spin-up (spin-down) electrons are shifted to $E=+M_z$ ($-M_z$), respectively. This gives rise to a pair of gapless dispersive edge modes crossing at band center $E=0$ as shown the blue lines in the inset of Fig.~\ref{fig1}(b). The RSOC admixes the spin-up and spin-down components, opening an edge-state gap around $E=0$  [see red lines in the inset of Fig.~\ref{fig1}(b)].

To verify the topological origin of the edge modes, we calculate the Chern number (CN) of the system using the non-commutative Kubo formula $C = 2\pi i \Tr[Q[\partial_{k_x} Q,\partial_{k_y}Q]]$, considering the periodic boundary conditions (PBC) in the $y$ and $x$ directions. Here $Q$ is the projector onto the occupied states of $H$. Similarly, the spin Chern numbers (SCNs) of spin-up and spin-down subsystems can be expressed as $C_{\pm} = 2\pi i\mathrm{Tr}[Q_{\pm}[\partial_{k_x} Q_{\pm},\partial_{k_y}Q_{\pm}]]$ \cite{PordanPRB2009,Prodan2011}, where $Q_{\pm}$ is spectral projector onto the positive/negative eigenvalue of $Q\sigma_zQ$. In Fig.~\ref{fig1}(c), we plot the the Chern numbers as a function of $E_F$.
When the chemical potential lies within the energy gap of the inverted zeroth Landau levels $E_F\in[-0.14,0.14]$, the Chern number $C=0$, while the spin Chern numbers $C_{\pm}=\pm1$~[Fig.~\ref{fig1}(c)] resembling those observed in the QSHE.
This indicates that helical the edge modes originate from the nonzero spin Chern numbers and are protected by $CT$ invariance, where $C$ represents the charge conjugation operation and $T$ is the time-reversal operation~\cite{QFSunPRL2010}.


\section{Effect of disorder and temperature on edge transport in the presence of RSOC}

Within the Laudauer-B$\ddot{\rm{u}}$ttiker transport formalism, the transmission coefficient from the terminal $q$ to the terminal $p$ can be obtained from $T_{pq}(E_F)=$Tr$[\Gamma_pG^r\Gamma_qG^a]$, where  $G^r=[G^a]^{\dagger}$ represents the Green's function and the line width function $\Gamma_p=i[\Sigma_p^r-\Sigma_p^a]$ \cite{datta1995}. The current in lead $p$ is given by $I_p=\sum_{q}\tilde{T}_{pq}(T,E_F)(V_p-V_q)$ with $V_p$ is the bias in the lead $p$. Here, the transmission coefficient at finite temperatures $T$ is given by $\tilde{T}_{pq}(T,E_F)=\int T_{pq}(\epsilon)(-\frac{\partial f_0}{\partial \epsilon})\mathrm{d}\epsilon$~\cite{datta1995}, where the Fermi distribution $f_0=[e^{(\epsilon-E_{F})/k_BT}+1]^{-1}$ and $k_B$ is the Boltzmann constant.
When a current $I_{14}$ flowing from lead 1 to lead 4,
 longitudinal conductance $G_{14,23}=G_{xx}=I_{14}/(V_2-V_3)$,  where $I_{14}=-I_1=I_4$.


In Fig.~\ref{fig1}(d), we show the conductance $G_{xx}$ as a function of the chemical potential $E_F$.
In the absence of RSOC, we observe that $G_{xx}=2e^2/h$ when $E_F\in[-0.14,0.14]$ (see the blue line).
The conductance plateau arises from the ballistic transport of the gapless helical channel modes [the blue lines in the inset of Fig.~\ref{fig1}(b)].
In contrast, when $\lambda_R=0.005$,  the longitudinal conductance $G_{xx}$ exhibits a pronounced dip at $E_F=0$ (see the red line), attributed the edge gap opened by RSOC~[see the red lines in Fig.~\ref{fig1}(b)]. 

Disorder usually plays a crucial role in the quantum Hall effect \cite{RevModPhys.67.357}. To investigate the impact of disorder on the helical quantum Hall phase,
we introduce the disorder term as $H_w = \sum_i c_i^{\dagger}V_i  c_i$, where $V_i=U_i \sigma_0$ represents non-magnetic disorder, or $V_i=U_i \sigma_{i=x,y,z}$ represents the magnetic disorder. Here $\sigma_0$ denotes the $2\times2$ identity matrix and $U_i$ is uniformly distributed in the range $ [-W/2, W/2]$ with $W$ representing the disorder strength. For nonmagnetic disorder, when $\lambda_R=0$, the longitudinal conductance $G_{xx}$ at $E_F=0$ remain quantized as long as $W<1.8$ ~[Fig.~\ref{fig1}(e) dashed black line]. However, when $W>1.8$, the helical edge modes are scattered into the zeroth Landau level at $E=\pm0.15$, leading to the suppression of ballistic transport. This observation suggests that the helical edge modes in the helical quantum Hall phase exhibit similar robust transport characteristics to those observed in the QSHE.
Interestingly, when $\lambda_R=0.005$, $G_{xx}$ at $E_F=0$ gradually increases from $1.1e^2/h$ to $1.8e^2/h$ at zero temperature [the black line in Fig.~\ref{fig1}(e)] as the disorder strength increase from $W=0$ to $1.1$. This disorder-enhanced edge transport persists until $k_BT=0.01$ [the blue line in Fig.~\ref{fig1}(e)]. These findings indicate that weak nonmagnetic disorder can enhance the edge conductance in the low-temperature regime.

On the other hand, when examining a clean sample ($W=0$) at $E_F=0$, it is observed that the conductance tends to reach $2e^2/h$ as the temperature increase, as shown by the different colored lines in Fig.~\ref{fig1}(e). The behavior of the finite-temperature conductance $G_{xx}(E_F=0,T)$ at $E_F=0$ can be approximated by considering the finite-temperature transmission coefficient $\tilde{T}_{pq}(T,E_F)=\int T_{pq}(\epsilon)(-\frac{\partial f_0}{\partial \epsilon})\mathrm{d}\epsilon$, where $-\frac{\partial f_0}{\partial \epsilon}$ represents the thermal broadening function. 
Consequently, $G_{xx}(E_F=0,T)$ at $E_F=0$ can be estimated as a weighted average of zero-temperature conductance $G_{xx}(E_F,T=0)$ at various $E_F$. As the temperature increases, the contribution of conductance away from $E_F=0$ becomes more significant, leading to an increase in $G_{xx}(E_F=0,T)$.
In summary, we calculate $G_{xx}$ in the band gap induced by the RSOC for various $W$ and $T$ in Fig.~\ref{fig1}(f). The results demonstrate that weak disorder can enhance the longitudinal conductance $G_{xx}$ at low temperature. Additionally, for $0\leq W<1.2$, the conductance $G_{xx}$ can also be increased by raising temperature.

\begin{figure}[tbh]
	\centering
	\includegraphics[width=3.4in]{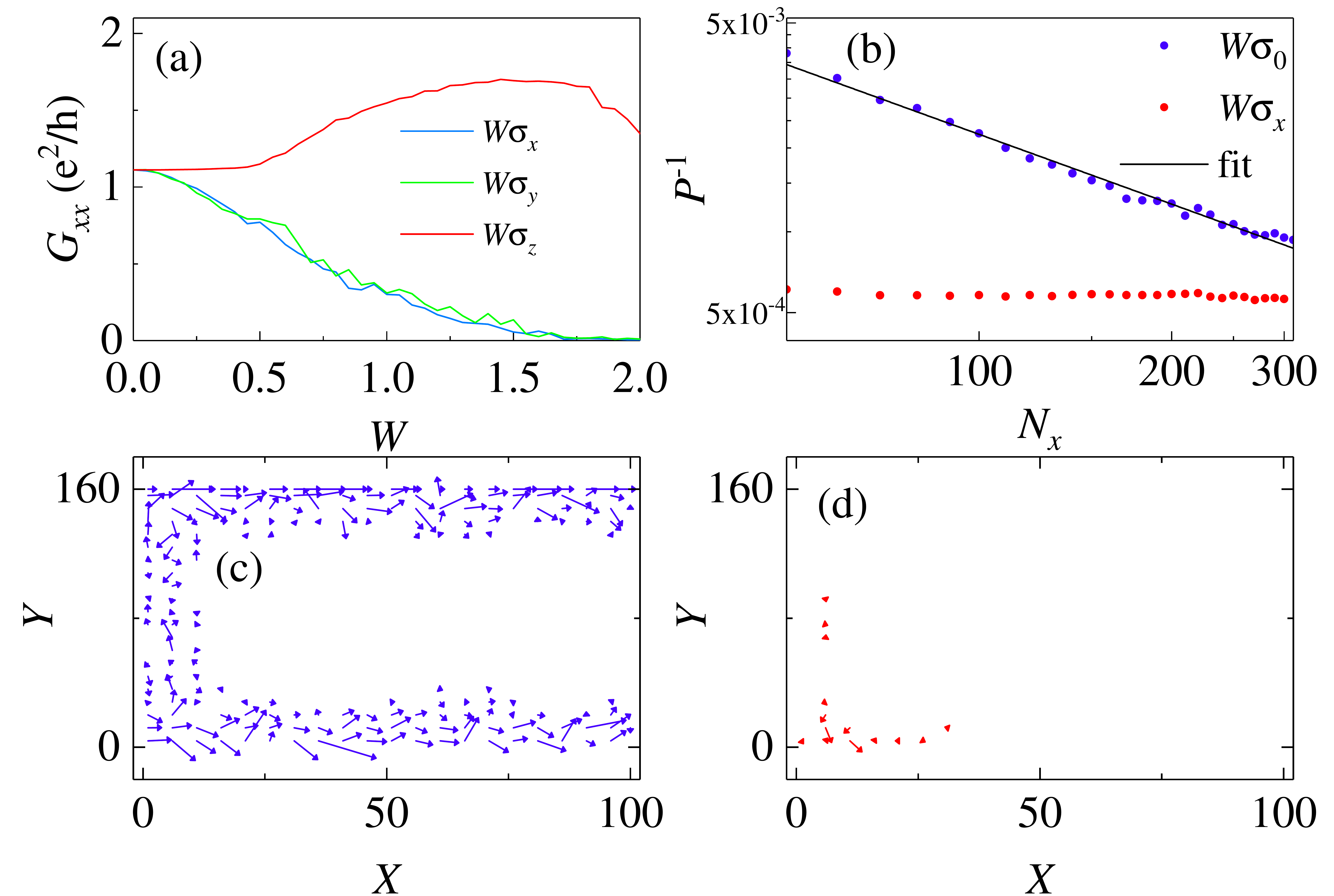}
	\caption{(Color online). 
		(a) $G_{xx}$ versus $W$ for different types of disorder.
		(b) IPR $P^{-1}$ as a function of number of sites in $x$ direction $N_x$ with strength $W=1.5$. $\sigma_0$-type and $\sigma_x$-type disorder are marked by blue and red points, and the linear fit of blue points is shown as black line.
		The local-current-flow configurations for (c) $\sigma_0$-type and (d) $\sigma_x$-type disorder with $W=1.5$ and size $N_y\times N_x=160\times 100$.
  The data in  (a) and (b) are averaged over $200$ and $8000$ random configurations, respectively. Other parameters are the same as those in Fig. \ref{fig1}.
		\label{fig2} }
\end{figure}

For comparison purposes, we plot $G_{xx}$ for different types of magnetic disorder in Fig.~\ref{fig2}(a). It can be observed that both $\sigma_x$-type and $\sigma_y$-type disorder have a similar effect on  $G_{xx}$, resulting in a monotonic suppression of $G_{xx}$.
This can be attributed to the coupling between the spin-up and spin-down helical edge modes caused by  $\sigma_x$-type and $\sigma_y$-type disorder, which leads to strong backscattering. In contrast, weak $\sigma_z$-type disorder, which conserves spin in the $z$ direction, actually enhances $G_{xx}$. However, it is important to note that the enhancement caused by $\sigma_z$-type disorder is weaker compared to nonmagnetic disorder.

To provide a clearer distinction between $\sigma_0$-type and $\sigma_x$-type disorder, we calculate he inverse participation ratio (IPR) and the local-current-flow vector at a moderate disorder strength of $W=1.5$.
Figure~\ref{fig2}(b) depicts IPR as a function of number of sites in $x$ direction $N_x$. IPR reflects the spatial extension of the eigenstates and the average IPR is defined as \cite{JANSSEN19981,PJH2015,Wegner1980,ZYY2009}
\begin{equation*}
P^{-1}=\Bigg\langle\frac{\sum_i^{N_{t}}\mid\psi_i\mid^4}{[\sum_i^{N_t}\mid\psi_i\mid^2]^2}\Bigg\rangle, \nonumber
\end{equation*}
where $N_t=N_y\times N_x$ is the total number of sites, $\psi_i$ is the wave function of site $i$, and $\langle...\rangle$ represents the disorder average. In the thermodynamic limit, IPR $P^{-1}$ approaches constant for localized states, while it scales as $P^{-1}\propto 1/N_x$ for extended edge states \cite{JANSSEN19981,JTEdwards1972,LHL2021}. As illustrated in Fig.~\ref{fig2}(b), the blue points exhibit a well-fitted linear relationship [indicated by the black line], indicating the extended nature of the wave functions under $\sigma_0$-type disorder. However, if $\sigma_x$-type disorder is applied, $P^{-1}$ is independent of sample length $N_x$ [Fig.~\ref{fig2}(b) red points], indicating that the states are localized due to the presence of disorder.
Moreover, we plot Figs.~\ref{fig2}(c)-\ref{fig2}(d)  the local currents \cite{HuaJiangPRB2009,ZZQPRB2019,ZZQPRB2021} observed in a device with dimensions $N_y\times N_x=160\times100$ under $\sigma_0$-type and $\sigma_x$-type disorder, respectively.
For $\sigma_0$-type disorder, the local currents locate mainly on the $y$ directional edge [Fig.~\ref{fig2}(c)], while the local currents are localized by $\sigma_x$-type disorder as shown in Fig.~\ref{fig2}(d).

\begin{figure}[tbh]
	\centering
	\includegraphics[width=3.4in]{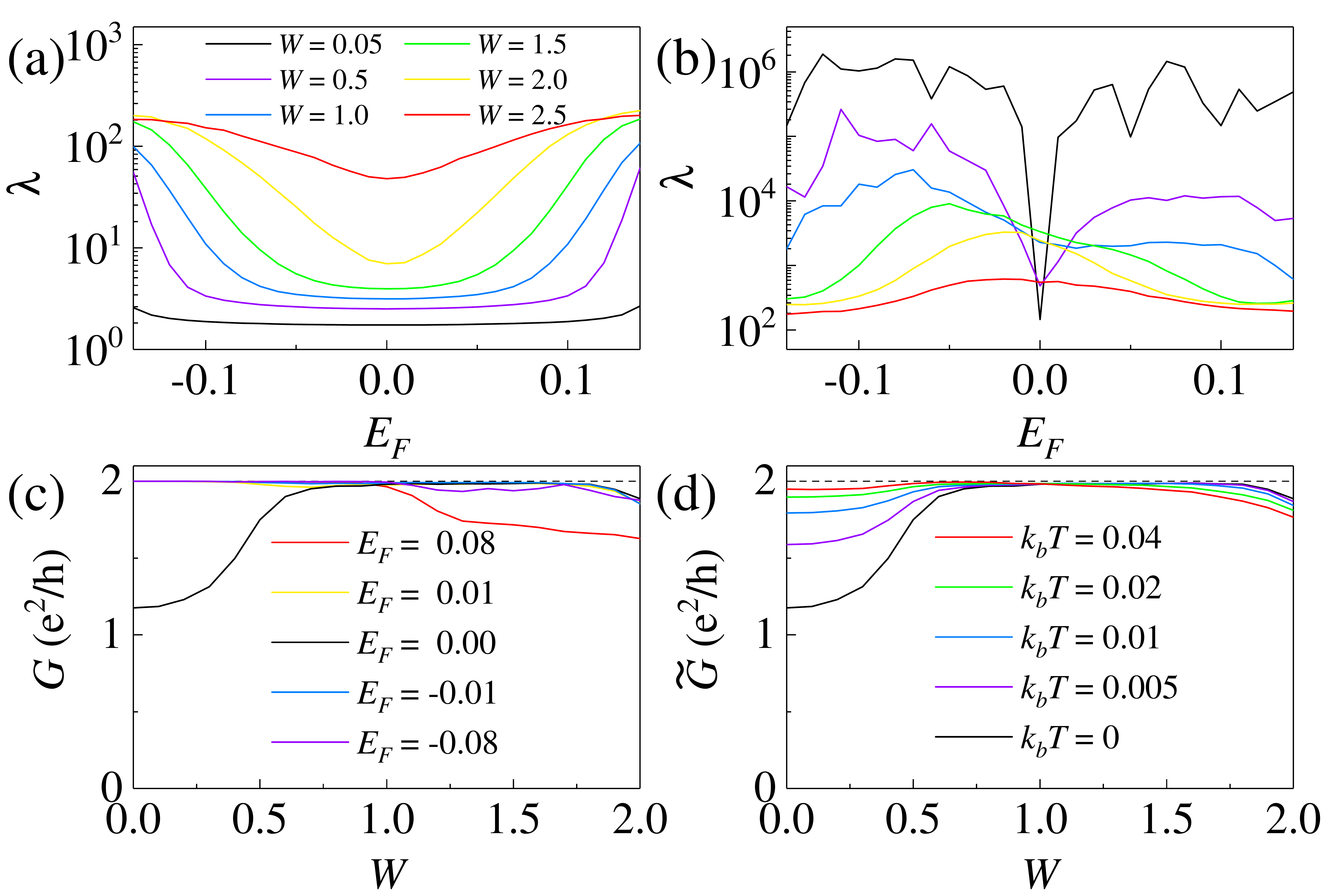}
	\caption{(Color online). Localization length $\lambda$ as a function of chemical potential $E_F$ for various disorder strength $W$ under (a) PBC and (b) OBC, respectively,  $\phi=\pi/40$. (c) Conductance $G$ calculated by the transfer matrix method versus $W$ for different chemical potential $E_F$. (d) $G$ versus $W$ for various temperatures $k_BT$. Other parameters $N=320$ ($N_y=160$), $L=100$, $\phi=0.05$ and $M_z=0.15$.
		\label{fig3} }
\end{figure}

\section{Origin of the disorder-enhanced conductance}
To elucidate the underlying cause of the disorder-enhanced conductance, we proceed with the calculation of the localization length using the transfer matrix method \cite{MacKinnon1981,MacKinnon1983,Kramer1993}. We consider a 2D finite  sample with dimensions $N\times L$, where $N$ represents the number of propagating channels along the length $L$. Utilizing the transfer matrix method, we determine the Lyapunov exponents (LEs), denoted as $\gamma_i$, which characterize the exponential decay behavior of the corresponding channels. LEs can be obtained by calculating the logarithms of the positive eigenvalues of the transfer matrix.
The localization length $\lambda$ and conductance $G$ without leads can be respectively expressed in terms of LEs as follows:
\begin{eqnarray}
\lambda=\frac{1}{\gamma_{\mathrm{min}}}, \;\;\;\; G=\sum_i^N\frac{1}{\cosh^2(\gamma_iL)},
\label{Eq.2}
\end{eqnarray}
 where $\gamma_{\mathrm{min}}$ is the smallest LE \cite{Beenakker1997,SJXiong2007}.
Note that to ensure stability of the numerical calculation,
a negligible term representing the next-nearest-neighbor hopping term $-\sum_{\langle\langle ij\rangle\rangle,\sigma}t_2e^{i\phi_{ij}}c_{i\sigma}^{\dagger}c_{j\sigma}$ is added to the model Hamiltonian with $t_2=0.005$. This additional term only shift the gap center $E_0$ from $0$ to $-0.015$, and its influence on the energy band structure can be ignored.
As shown in Fig.~\ref{fig3}(a), employing the PBC in the $y$ direction results in small the renormalized localization lengths ($\lambda/N_y<1$), indicating an insulating behavior.
On the contrary, when employing the OBC as shown in Fig.~\ref{fig3}(b), $\lambda$
near $E_F=0$ can be significantly enhanced either by introducing weak disorder or  by shifting the Fermi energy outside the  gap of edge states. The observed enhancement of $\lambda$ by transitioning from the PBC to the OBC strongly suggests the dominant role played by the helical edge modes in determining the conductance. According to Eq.~(\ref{Eq.2}), for large $\lambda$, $\gamma_{\mathrm{min}}\to0^+$, the contribution of corresponding channel to conductance $G$ tends to quantized value $e^2/h$ (considering two-spin species, $G\to2e^2/h$). Similarly, the conductance at finite temperature can be expressed as $\tilde{G}(T,E_F)=\int G(\epsilon)(-\frac{\partial f_0}{\partial \epsilon})\mathrm{d}\epsilon$. In Figs.~\ref{fig3}(c)-\ref{fig3}(d), we calculate $G$ as a function of $W$ for various $E_F$ and temperatures $T$. By applying moderate disorder, the conductance $G$ at $E_F=0$ can be enhanced [see Figs.~\ref{fig3}(c)-\ref{fig3}(d)] due to the increased localization length. Additionally, raising the temperature also amplifies the conductance $\tilde{G}(E_F=0)$ as illustrated in Fig.~\ref{fig3}(d),
since $\tilde{G}(T,E_F=0)$ is a weighted average of zero temperature conductance $G(E_F)$ at different $E_F$.
It is worth mentioning that the conductance $G$ calculated using transfer matrix method is independent of leads, excluding the influence of contact resistance \cite{YYZhangPRL2009}. Consequently, $G$ is slightly larger than $G_{xx}$~[Figs.~\ref{fig1}(e) and (f)].

\begin{figure}[tbh]
	\centering
	\includegraphics[width=3.4in]{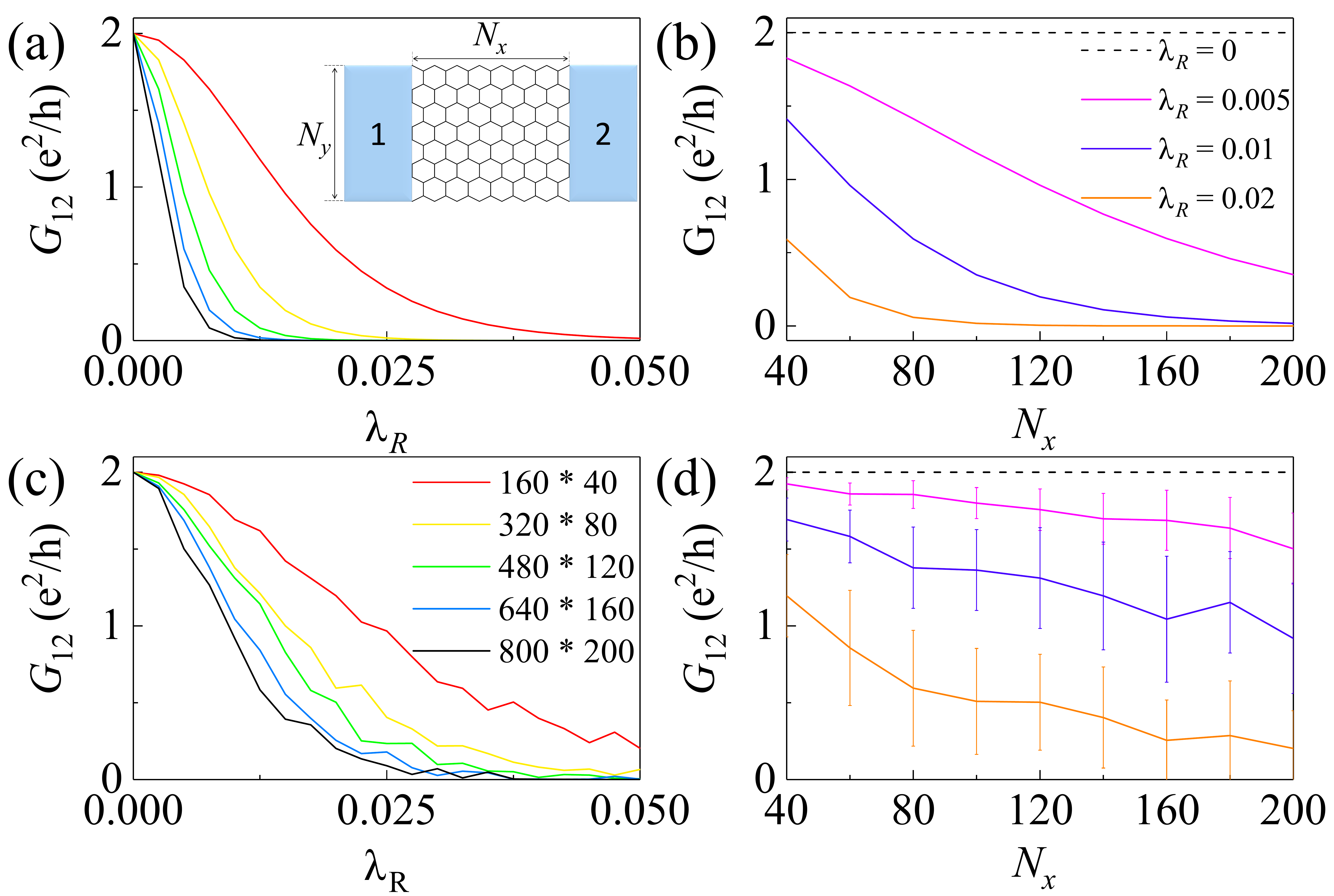}
	\caption{(Color online). Two-terminal conductance $G_{12}$ versus RSOC $\lambda_R$ for different system sizes $N_y\times N_x$ with (a) disorder strength $W=0$ and (c) $W=1$, respectively. $G_{12}$ as a function of system size for various $\lambda_R$ with (b) $W=0$ and (d) $W=1$, respectively. The error bar in (d) denotes conductance fluctuation. The inset of (a) shows a two-terminal device. $G_{12}$ are averaged over $48$ random configurations. $E_F=0$, other parameters are the same as those in Fig. \ref{fig1}.
		\label{fig4} }
\end{figure}

\section{RSOC effect on longitudinal conductance} Furthermore, we conduct further investigations into edge transport in the presence of varying RSOC by examining the longitudinal conductance $G_{12}$ in a two-terminal device~[inset of Fig.~\ref{fig4}(a)] \cite{HuaJiangPRB2009}. The dependence of $G_{12}$ on the RSOC $\lambda_R$ and system size $N_y\times N_x$ of clean system ($W=0$) are plotted in Figs.~\ref{fig4}(a) and \ref{fig4}(b), while the case of $W=1$ are shown in Figs.~\ref{fig4}(c) and \ref{fig4}(d). When $\lambda_R=0$, the conductance exhibits a quantized value of $G_{12}=2e^2/h$, which is independent of system size. As $\lambda_R$ increases, the coupling between the spin-up and spin-down channels of helical edge states becomes stronger leading to increased backscattering between them and consequently reducing the conductance. For a given $\lambda_R>0$, as the size increases, the probability of backscattering among edge modes also increases, resulting in a monotonic decrease in $G_{12}$. Applying weak disorder slows down the decrease in conductance, as demonstrated in Figs.~\ref{fig4}(a) and \ref{fig4}(c) for varying $\lambda_R$ and Figs.~\ref{fig4}(b) and \ref{fig4}(d) for different system sizes.

\begin{figure}[tbh]
	\centering
	\includegraphics[width=3.4in]{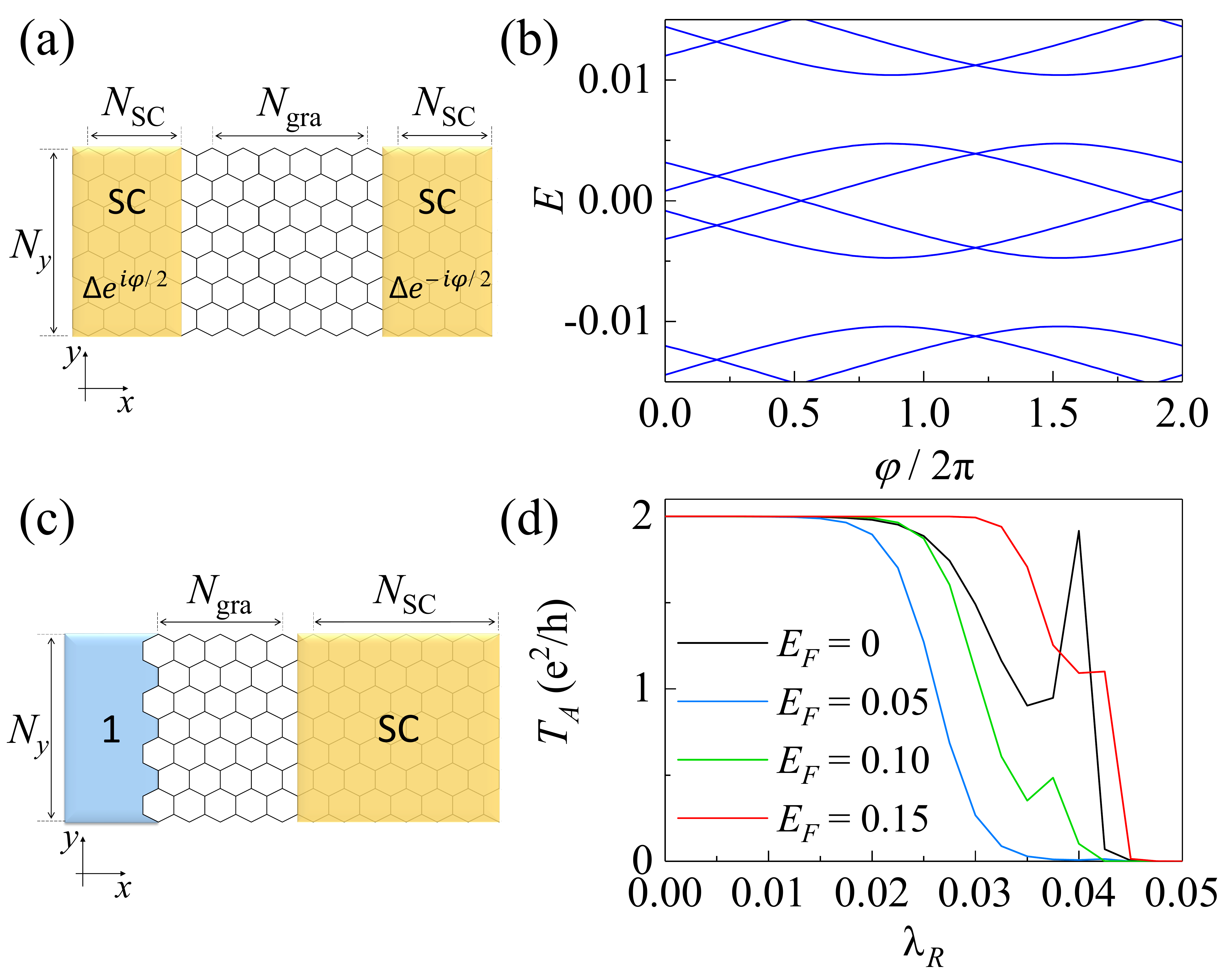}
	\caption{(Color online). (a) Schematic picture of a SC/graphene/SC junction, where $\varphi$ is the phase difference between two SC regions, and $N_y=320$, $N_{\mathrm{gra}}=100$, $N_{\mathrm{SC}}=50$. (b) Andreev bound state spectrum of the Josephson junction as a function of $\varphi$ with $E_F=0$. (c) Schematic diagram for a lead/graphene/SC device. Here $N_y=320$, $N_{\mathrm{gra}}=50$ and $N_{\mathrm{SC}}=100$. (d) Andreev reflection $T_A$ at zero energy from the lead versus RSOC $\lambda_R$ for various chemical potential $E_F$. For the SC region, the superconductor gap $\Delta=0.05$, $\phi=0$ and $M_z=0$. The parameters of graphene region are the same as those in Fig. \ref{fig1}.
		\label{fig5} }
\end{figure}

\section{ Superconductivity and Majorana zero modes} In proximity to an $s$-wave superconductor (SC), the helical quantum Hall phase of the graphene serves as a natural platform for hosting Majorana zero modes [as shown in Fig.~\ref{fig5}(a)]. The effective Hamiltonian of graphene and SC heterostructure can be written as \cite{QFSunJPCM2009}
\begin{eqnarray}
	H_{GS}=H +\sum_i (\Delta_{i} c_{i\uparrow}^{\dagger}c_{i\downarrow}^{\dagger}+\mathrm{H}.\mathrm{c}.),\nonumber
\end{eqnarray}
where $\Delta_{i}$ is the superconducting paring potential at the site $i$.
 Due to the Meissner effect, the magnetic field is zero within the SC region, and thus we set $\phi=0$ and $M_z=0$ in this region.


As depicted in Fig.~\ref{fig5}(a), the Josephson junction can be formed by coupling two $s$-wave SCs to the two ends of graphene, where $\Delta e^{\pm i\varphi/2}$ denotes the pairing order parameter of the SCs with a tunable phase $\varphi$ \cite{CZChenPRB2018,CZChenPRB20182}. The presence of SCs induces a superconducting gap on the helical edge modes, resulting in the creation of a small number of Andreev bound states within the gap.
Figure~\ref{fig5}(b) shows the Andreev bound state spectrum as a function of the phase difference $\varphi$, the bands crossing at zero energy signals the emergence of Majorana zero modes.

To explain the origin of the Andreev bound states, we consider a single lead coupled to the grephene/SC heterostructure [see Fig.~\ref{fig5}(a)] to investigate the Andreev reflection (AR) process at the SC and graphene interface. AR involves an incident electron from the metal state being reflected as a hole, thereby generating a Cooper pair within the superconductor. The  AR coefficient can be evaluated by using the Green's functions $T_A=\mathrm{Tr}[\Gamma_{Le}G_{eh}^r\Gamma_{Lh}G_{he}^a]$, where $\Gamma_{Le(h)}$ is the electron (hole)'s line width function in the left lead,
and $G_{eh}^r$ is the full Green's function of the system \cite{QFSunJPCM2009,JieLiuPRB2014,JieLiuPRB2017}.

When the chemical potential $E_F$ lies within the gap of the helical quantum spin Hall, there is only on channel per edge. Due to the SC gap,
the probability of transmission from the lead to SC tends to $0$, while the probability of reflection tends to $1$.
As a result, for small $\lambda_R$, $T_A=2e^2/h$ as shown in Fig.~\ref{fig5}(d).
Therefore, we conclude that the zero-energy Andree bound states in Fig.~\ref{fig5}(b) arise from the quantized AR at the interface between the SC and graphene.
Moreover, the quantized AR at zero energy is also the fingerprint of the emergence of a MZM in the grephene/SC heterostructure region \cite{KTLawPRL2009,FidkowskiPRB2012,HaimPRL2019}.



\section{ Discussion and conclusion}
In conclusion, our study reveals that in a graphene nanoribbon subjected to a perpendicular magnetic field, the suppressed longitudinal conductance by RSOC can be enhanced by raising the temperature under weak disorder or by applying weak nonmagnetic Anderson disorder at low temperatures. Remarkably, we demonstrate that this disorder increased conductance can approach the quantized value of $2e^2/h$, in certain regions. These findings align qualitatively with the transport results of the helical quantum Hall phase observed in the recent experiments~\cite{VeyratScience2020}.
Furthermore, we demonstrate that, in proximity to superconductivity, such a helical quantum Hall phase in graphene has the potential to host Majorana zero modes.

\section*{Acknowledgments}
We thank Zhi-Qiang Zhang for fruitful discussions. The authors acknowledge the support by the National Key R\&D Program of China (Grants No. 2022YFA1403700), and the NSFC (under Grants No. 12074108, No. 11974256, and No. 12147102), the Priority Academic Program Development (PAPD) of Jiangsu Higher Education Institution, and the Natural Science Foundation of Chongqing (Grant No. CSTB2022NSCQ-MSX0568).

\bibliographystyle{apsrev4-2} 
\bibliography{ref_qshhe}

\begin{thebibliography}{58}%
\makeatletter
\providecommand \@ifxundefined [1]{%
 \@ifx{#1\undefined}
}%
\providecommand \@ifnum [1]{%
 \ifnum #1\expandafter \@firstoftwo
 \else \expandafter \@secondoftwo
 \fi
}%
\providecommand \@ifx [1]{%
 \ifx #1\expandafter \@firstoftwo
 \else \expandafter \@secondoftwo
 \fi
}%
\providecommand \natexlab [1]{#1}%
\providecommand \enquote  [1]{``#1''}%
\providecommand \bibnamefont  [1]{#1}%
\providecommand \bibfnamefont [1]{#1}%
\providecommand \citenamefont [1]{#1}%
\providecommand \href@noop [0]{\@secondoftwo}%
\providecommand \href [0]{\begingroup \@sanitize@url \@href}%
\providecommand \@href[1]{\@@startlink{#1}\@@href}%
\providecommand \@@href[1]{\endgroup#1\@@endlink}%
\providecommand \@sanitize@url [0]{\catcode `\\12\catcode `\$12\catcode
  `\&12\catcode `\#12\catcode `\^12\catcode `\_12\catcode `\%12\relax}%
\providecommand \@@startlink[1]{}%
\providecommand \@@endlink[0]{}%
\providecommand \url  [0]{\begingroup\@sanitize@url \@url }%
\providecommand \@url [1]{\endgroup\@href {#1}{\urlprefix }}%
\providecommand \urlprefix  [0]{URL }%
\providecommand \Eprint [0]{\href }%
\providecommand \doibase [0]{https://doi.org/}%
\providecommand \selectlanguage [0]{\@gobble}%
\providecommand \bibinfo  [0]{\@secondoftwo}%
\providecommand \bibfield  [0]{\@secondoftwo}%
\providecommand \translation [1]{[#1]}%
\providecommand \BibitemOpen [0]{}%
\providecommand \bibitemStop [0]{}%
\providecommand \bibitemNoStop [0]{.\EOS\space}%
\providecommand \EOS [0]{\spacefactor3000\relax}%
\providecommand \BibitemShut  [1]{\csname bibitem#1\endcsname}%
\let\auto@bib@innerbib\@empty
\bibitem [{\citenamefont {Kane}\ and\ \citenamefont
  {Mele}(2005{\natexlab{a}})}]{KaneMelePRL20051}%
  \BibitemOpen
  \bibfield  {author} {\bibinfo {author} {\bibfnamefont {C.~L.}\ \bibnamefont
  {Kane}}\ and\ \bibinfo {author} {\bibfnamefont {E.~J.}\ \bibnamefont
  {Mele}},\ }\href {https://doi.org/10.1103/PhysRevLett.95.146802} {\bibfield
  {journal} {\bibinfo  {journal} {Phys. Rev. Lett.}\ }\textbf {\bibinfo
  {volume} {95}},\ \bibinfo {pages} {146802} (\bibinfo {year}
  {2005}{\natexlab{a}})}\BibitemShut {NoStop}%
\bibitem [{\citenamefont {Kane}\ and\ \citenamefont
  {Mele}(2005{\natexlab{b}})}]{KaneMelePRL20052}%
  \BibitemOpen
  \bibfield  {author} {\bibinfo {author} {\bibfnamefont {C.~L.}\ \bibnamefont
  {Kane}}\ and\ \bibinfo {author} {\bibfnamefont {E.~J.}\ \bibnamefont
  {Mele}},\ }\href {https://doi.org/10.1103/PhysRevLett.95.226801} {\bibfield
  {journal} {\bibinfo  {journal} {Phys. Rev. Lett.}\ }\textbf {\bibinfo
  {volume} {95}},\ \bibinfo {pages} {226801} (\bibinfo {year}
  {2005}{\natexlab{b}})}\BibitemShut {NoStop}%
\bibitem [{\citenamefont {Min}\ \emph {et~al.}(2006)\citenamefont {Min},
  \citenamefont {Hill}, \citenamefont {Sinitsyn}, \citenamefont {Sahu},
  \citenamefont {Kleinman},\ and\ \citenamefont {MacDonald}}]{HKMinPRB2006}%
  \BibitemOpen
  \bibfield  {author} {\bibinfo {author} {\bibfnamefont {H.}~\bibnamefont
  {Min}}, \bibinfo {author} {\bibfnamefont {J.~E.}\ \bibnamefont {Hill}},
  \bibinfo {author} {\bibfnamefont {N.~A.}\ \bibnamefont {Sinitsyn}}, \bibinfo
  {author} {\bibfnamefont {B.~R.}\ \bibnamefont {Sahu}}, \bibinfo {author}
  {\bibfnamefont {L.}~\bibnamefont {Kleinman}},\ and\ \bibinfo {author}
  {\bibfnamefont {A.~H.}\ \bibnamefont {MacDonald}},\ }\href
  {https://doi.org/10.1103/PhysRevB.74.165310} {\bibfield  {journal} {\bibinfo
  {journal} {Phys. Rev. B}\ }\textbf {\bibinfo {volume} {74}},\ \bibinfo
  {pages} {165310} (\bibinfo {year} {2006})}\BibitemShut {NoStop}%
\bibitem [{\citenamefont {Yao}\ \emph {et~al.}(2007)\citenamefont {Yao},
  \citenamefont {Ye}, \citenamefont {Qi}, \citenamefont {Zhang},\ and\
  \citenamefont {Fang}}]{YYgPRB2007}%
  \BibitemOpen
  \bibfield  {author} {\bibinfo {author} {\bibfnamefont {Y.}~\bibnamefont
  {Yao}}, \bibinfo {author} {\bibfnamefont {F.}~\bibnamefont {Ye}}, \bibinfo
  {author} {\bibfnamefont {X.-L.}\ \bibnamefont {Qi}}, \bibinfo {author}
  {\bibfnamefont {S.-C.}\ \bibnamefont {Zhang}},\ and\ \bibinfo {author}
  {\bibfnamefont {Z.}~\bibnamefont {Fang}},\ }\href
  {https://doi.org/10.1103/PhysRevB.75.041401} {\bibfield  {journal} {\bibinfo
  {journal} {Phys. Rev. B}\ }\textbf {\bibinfo {volume} {75}},\ \bibinfo
  {pages} {041401} (\bibinfo {year} {2007})}\BibitemShut {NoStop}%
\bibitem [{\citenamefont {Sheng}\ \emph {et~al.}(2005)\citenamefont {Sheng},
  \citenamefont {Sheng}, \citenamefont {Ting},\ and\ \citenamefont
  {Haldane}}]{ShengLPRL2005}%
  \BibitemOpen
  \bibfield  {author} {\bibinfo {author} {\bibfnamefont {L.}~\bibnamefont
  {Sheng}}, \bibinfo {author} {\bibfnamefont {D.~N.}\ \bibnamefont {Sheng}},
  \bibinfo {author} {\bibfnamefont {C.~S.}\ \bibnamefont {Ting}},\ and\
  \bibinfo {author} {\bibfnamefont {F.~D.~M.}\ \bibnamefont {Haldane}},\ }\href
  {https://doi.org/10.1103/PhysRevLett.95.136602} {\bibfield  {journal}
  {\bibinfo  {journal} {Phys. Rev. Lett.}\ }\textbf {\bibinfo {volume} {95}},\
  \bibinfo {pages} {136602} (\bibinfo {year} {2005})}\BibitemShut {NoStop}%
\bibitem [{\citenamefont {Bernevig}\ and\ \citenamefont
  {Zhang}(2006)}]{BernevigPRL2006}%
  \BibitemOpen
  \bibfield  {author} {\bibinfo {author} {\bibfnamefont {B.~A.}\ \bibnamefont
  {Bernevig}}\ and\ \bibinfo {author} {\bibfnamefont {S.-C.}\ \bibnamefont
  {Zhang}},\ }\href {https://doi.org/10.1103/PhysRevLett.96.106802} {\bibfield
  {journal} {\bibinfo  {journal} {Phys. Rev. Lett.}\ }\textbf {\bibinfo
  {volume} {96}},\ \bibinfo {pages} {106802} (\bibinfo {year}
  {2006})}\BibitemShut {NoStop}%
\bibitem [{\citenamefont {Murakami}(2006)}]{NurakamiPRL2006}%
  \BibitemOpen
  \bibfield  {author} {\bibinfo {author} {\bibfnamefont {S.}~\bibnamefont
  {Murakami}},\ }\href {https://doi.org/10.1103/PhysRevLett.97.236805}
  {\bibfield  {journal} {\bibinfo  {journal} {Phys. Rev. Lett.}\ }\textbf
  {\bibinfo {volume} {97}},\ \bibinfo {pages} {236805} (\bibinfo {year}
  {2006})}\BibitemShut {NoStop}%
\bibitem [{\citenamefont {Fu}\ \emph {et~al.}(2007)\citenamefont {Fu},
  \citenamefont {Kane},\ and\ \citenamefont {Mele}}]{FuLiangPRL2007}%
  \BibitemOpen
  \bibfield  {author} {\bibinfo {author} {\bibfnamefont {L.}~\bibnamefont
  {Fu}}, \bibinfo {author} {\bibfnamefont {C.~L.}\ \bibnamefont {Kane}},\ and\
  \bibinfo {author} {\bibfnamefont {E.~J.}\ \bibnamefont {Mele}},\ }\href
  {https://doi.org/10.1103/PhysRevLett.98.106803} {\bibfield  {journal}
  {\bibinfo  {journal} {Phys. Rev. Lett.}\ }\textbf {\bibinfo {volume} {98}},\
  \bibinfo {pages} {106803} (\bibinfo {year} {2007})}\BibitemShut {NoStop}%
\bibitem [{\citenamefont {Liu}\ \emph {et~al.}(2008)\citenamefont {Liu},
  \citenamefont {Hughes}, \citenamefont {Qi}, \citenamefont {Wang},\ and\
  \citenamefont {Zhang}}]{LiuChaoxingPRL2008}%
  \BibitemOpen
  \bibfield  {author} {\bibinfo {author} {\bibfnamefont {C.}~\bibnamefont
  {Liu}}, \bibinfo {author} {\bibfnamefont {T.~L.}\ \bibnamefont {Hughes}},
  \bibinfo {author} {\bibfnamefont {X.-L.}\ \bibnamefont {Qi}}, \bibinfo
  {author} {\bibfnamefont {K.}~\bibnamefont {Wang}},\ and\ \bibinfo {author}
  {\bibfnamefont {S.-C.}\ \bibnamefont {Zhang}},\ }\href
  {https://doi.org/10.1103/PhysRevLett.100.236601} {\bibfield  {journal}
  {\bibinfo  {journal} {Phys. Rev. Lett.}\ }\textbf {\bibinfo {volume} {100}},\
  \bibinfo {pages} {236601} (\bibinfo {year} {2008})}\BibitemShut {NoStop}%
\bibitem [{\citenamefont {Hsieh}\ \emph {et~al.}(2008)\citenamefont {Hsieh},
  \citenamefont {Qian}, \citenamefont {Wray}, \citenamefont {Xia},
  \citenamefont {Hor}, \citenamefont {Cava},\ and\ \citenamefont
  {Hasan}}]{HsiehDN2008}%
  \BibitemOpen
  \bibfield  {author} {\bibinfo {author} {\bibfnamefont {D.}~\bibnamefont
  {Hsieh}}, \bibinfo {author} {\bibfnamefont {D.}~\bibnamefont {Qian}},
  \bibinfo {author} {\bibfnamefont {L.}~\bibnamefont {Wray}}, \bibinfo {author}
  {\bibfnamefont {Y.}~\bibnamefont {Xia}}, \bibinfo {author} {\bibfnamefont
  {Y.}~\bibnamefont {Hor}}, \bibinfo {author} {\bibfnamefont {R.}~\bibnamefont
  {Cava}},\ and\ \bibinfo {author} {\bibfnamefont {M.~Z.}\ \bibnamefont
  {Hasan}},\ }\href {https://doi.org/10.1038/nature06843} {\bibfield  {journal}
  {\bibinfo  {journal} {Nature}\ }\textbf {\bibinfo {volume} {452}},\ \bibinfo
  {pages} {970} (\bibinfo {year} {2008})}\BibitemShut {NoStop}%
\bibitem [{\citenamefont {König}\ \emph {et~al.}(2007)\citenamefont {König},
  \citenamefont {Wiedmann}, \citenamefont {Brüne}, \citenamefont {Roth},
  \citenamefont {Buhmann}, \citenamefont {Molenkamp}, \citenamefont {Qi},\ and\
  \citenamefont {Zhang}}]{KonigS2007}%
  \BibitemOpen
  \bibfield  {author} {\bibinfo {author} {\bibfnamefont {M.}~\bibnamefont
  {König}}, \bibinfo {author} {\bibfnamefont {S.}~\bibnamefont {Wiedmann}},
  \bibinfo {author} {\bibfnamefont {C.}~\bibnamefont {Brüne}}, \bibinfo
  {author} {\bibfnamefont {A.}~\bibnamefont {Roth}}, \bibinfo {author}
  {\bibfnamefont {H.}~\bibnamefont {Buhmann}}, \bibinfo {author} {\bibfnamefont
  {L.}~\bibnamefont {Molenkamp}}, \bibinfo {author} {\bibfnamefont {X.-L.}\
  \bibnamefont {Qi}},\ and\ \bibinfo {author} {\bibfnamefont {S.-C.}\
  \bibnamefont {Zhang}},\ }\href {https://doi.org/10.1126/science.1148047}
  {\bibfield  {journal} {\bibinfo  {journal} {Science}\ }\textbf {\bibinfo
  {volume} {318}},\ \bibinfo {pages} {766} (\bibinfo {year}
  {2007})}\BibitemShut {NoStop}%
\bibitem [{\citenamefont {Knez}\ \emph {et~al.}(2011)\citenamefont {Knez},
  \citenamefont {Du},\ and\ \citenamefont {Sullivan}}]{KnezPRL2011}%
  \BibitemOpen
  \bibfield  {author} {\bibinfo {author} {\bibfnamefont {I.}~\bibnamefont
  {Knez}}, \bibinfo {author} {\bibfnamefont {R.-R.}\ \bibnamefont {Du}},\ and\
  \bibinfo {author} {\bibfnamefont {G.}~\bibnamefont {Sullivan}},\ }\href
  {https://doi.org/10.1103/PhysRevLett.107.136603} {\bibfield  {journal}
  {\bibinfo  {journal} {Phys. Rev. Lett.}\ }\textbf {\bibinfo {volume} {107}},\
  \bibinfo {pages} {136603} (\bibinfo {year} {2011})}\BibitemShut {NoStop}%
\bibitem [{\citenamefont {Fei}\ \emph {et~al.}(2017)\citenamefont {Fei},
  \citenamefont {Palomaki}, \citenamefont {Wu}, \citenamefont {Zhao},
  \citenamefont {Cai}, \citenamefont {Sun}, \citenamefont {Nguyen},
  \citenamefont {Finney}, \citenamefont {Xu},\ and\ \citenamefont
  {Cobden}}]{FeiZNP2017}%
  \BibitemOpen
  \bibfield  {author} {\bibinfo {author} {\bibfnamefont {Z.}~\bibnamefont
  {Fei}}, \bibinfo {author} {\bibfnamefont {T.}~\bibnamefont {Palomaki}},
  \bibinfo {author} {\bibfnamefont {S.}~\bibnamefont {Wu}}, \bibinfo {author}
  {\bibfnamefont {W.}~\bibnamefont {Zhao}}, \bibinfo {author} {\bibfnamefont
  {X.}~\bibnamefont {Cai}}, \bibinfo {author} {\bibfnamefont {B.}~\bibnamefont
  {Sun}}, \bibinfo {author} {\bibfnamefont {P.}~\bibnamefont {Nguyen}},
  \bibinfo {author} {\bibfnamefont {J.}~\bibnamefont {Finney}}, \bibinfo
  {author} {\bibfnamefont {X.}~\bibnamefont {Xu}},\ and\ \bibinfo {author}
  {\bibfnamefont {D.~H.}\ \bibnamefont {Cobden}},\ }\href
  {https://doi.org/https://doi.org/10.1038/nphys4091} {\bibfield  {journal}
  {\bibinfo  {journal} {Nature Phys}\ }\textbf {\bibinfo {volume} {13}},\
  \bibinfo {pages} {677} (\bibinfo {year} {2017})}\BibitemShut {NoStop}%
\bibitem [{\citenamefont {Hatsuda}\ \emph {et~al.}(2018)\citenamefont
  {Hatsuda}, \citenamefont {Mine}, \citenamefont {Nakamura}, \citenamefont
  {Li}, \citenamefont {Wu}, \citenamefont {Katsumoto},\ and\ \citenamefont
  {Haruyama}}]{HatsudaSA2018}%
  \BibitemOpen
  \bibfield  {author} {\bibinfo {author} {\bibfnamefont {K.}~\bibnamefont
  {Hatsuda}}, \bibinfo {author} {\bibfnamefont {H.}~\bibnamefont {Mine}},
  \bibinfo {author} {\bibfnamefont {T.}~\bibnamefont {Nakamura}}, \bibinfo
  {author} {\bibfnamefont {J.}~\bibnamefont {Li}}, \bibinfo {author}
  {\bibfnamefont {R.}~\bibnamefont {Wu}}, \bibinfo {author} {\bibfnamefont
  {S.}~\bibnamefont {Katsumoto}},\ and\ \bibinfo {author} {\bibfnamefont
  {J.}~\bibnamefont {Haruyama}},\ }\href
  {https://doi.org/10.1126/sciadv.aau6915} {\bibfield  {journal} {\bibinfo
  {journal} {Science Advances}\ }\textbf {\bibinfo {volume} {4}},\ \bibinfo
  {pages} {eaau6915} (\bibinfo {year} {2018})}\BibitemShut {NoStop}%
\bibitem [{\citenamefont {Wu}\ \emph {et~al.}(2018)\citenamefont {Wu},
  \citenamefont {Fatemi}, \citenamefont {Gibson}, \citenamefont {Watanabe},
  \citenamefont {Taniguchi}, \citenamefont {Cava},\ and\ \citenamefont
  {Jarillo-Herrero}}]{WSFS2018}%
  \BibitemOpen
  \bibfield  {author} {\bibinfo {author} {\bibfnamefont {S.}~\bibnamefont
  {Wu}}, \bibinfo {author} {\bibfnamefont {V.}~\bibnamefont {Fatemi}}, \bibinfo
  {author} {\bibfnamefont {Q.~D.}\ \bibnamefont {Gibson}}, \bibinfo {author}
  {\bibfnamefont {K.}~\bibnamefont {Watanabe}}, \bibinfo {author}
  {\bibfnamefont {T.}~\bibnamefont {Taniguchi}}, \bibinfo {author}
  {\bibfnamefont {R.~J.}\ \bibnamefont {Cava}},\ and\ \bibinfo {author}
  {\bibfnamefont {P.}~\bibnamefont {Jarillo-Herrero}},\ }\href
  {https://doi.org/10.1126/science.aan6003} {\bibfield  {journal} {\bibinfo
  {journal} {Science}\ }\textbf {\bibinfo {volume} {359}},\ \bibinfo {pages}
  {76} (\bibinfo {year} {2018})}\BibitemShut {NoStop}%
\bibitem [{\citenamefont {Novoselov}\ \emph {et~al.}(2004)\citenamefont
  {Novoselov}, \citenamefont {Geim}, \citenamefont {Morozov}, \citenamefont
  {Jiang}, \citenamefont {Zhang}, \citenamefont {Dubonos}, \citenamefont
  {Grigorieva},\ and\ \citenamefont {Firsov}}]{NovoselovS2004}%
  \BibitemOpen
  \bibfield  {author} {\bibinfo {author} {\bibfnamefont {K.~S.}\ \bibnamefont
  {Novoselov}}, \bibinfo {author} {\bibfnamefont {A.~K.}\ \bibnamefont {Geim}},
  \bibinfo {author} {\bibfnamefont {S.~V.}\ \bibnamefont {Morozov}}, \bibinfo
  {author} {\bibfnamefont {D.}~\bibnamefont {Jiang}}, \bibinfo {author}
  {\bibfnamefont {Y.}~\bibnamefont {Zhang}}, \bibinfo {author} {\bibfnamefont
  {S.~V.}\ \bibnamefont {Dubonos}}, \bibinfo {author} {\bibfnamefont {I.~V.}\
  \bibnamefont {Grigorieva}},\ and\ \bibinfo {author} {\bibfnamefont {A.~A.}\
  \bibnamefont {Firsov}},\ }\href {https://doi.org/10.1126/science.1102896}
  {\bibfield  {journal} {\bibinfo  {journal} {Science}\ }\textbf {\bibinfo
  {volume} {306}},\ \bibinfo {pages} {666} (\bibinfo {year}
  {2004})}\BibitemShut {NoStop}%
\bibitem [{\citenamefont {Novoselov}\ \emph {et~al.}(2005)\citenamefont
  {Novoselov}, \citenamefont {Geim}, \citenamefont {Morozov}, \citenamefont
  {Jiang}, \citenamefont {Katsnelson}, \citenamefont {Grigorieva},
  \citenamefont {Dubonos},\ and\ \citenamefont {Firsov}}]{NovoselovN2005}%
  \BibitemOpen
  \bibfield  {author} {\bibinfo {author} {\bibfnamefont {K.~S.}\ \bibnamefont
  {Novoselov}}, \bibinfo {author} {\bibfnamefont {A.~K.}\ \bibnamefont {Geim}},
  \bibinfo {author} {\bibfnamefont {S.~V.}\ \bibnamefont {Morozov}}, \bibinfo
  {author} {\bibfnamefont {D.}~\bibnamefont {Jiang}}, \bibinfo {author}
  {\bibfnamefont {M.~I.}\ \bibnamefont {Katsnelson}}, \bibinfo {author}
  {\bibfnamefont {I.~V.}\ \bibnamefont {Grigorieva}}, \bibinfo {author}
  {\bibfnamefont {S.~V.}\ \bibnamefont {Dubonos}},\ and\ \bibinfo {author}
  {\bibfnamefont {A.~A.}\ \bibnamefont {Firsov}},\ }\href
  {https://doi.org/10.1038/nature04233} {\bibfield  {journal} {\bibinfo
  {journal} {Nature}\ }\textbf {\bibinfo {volume} {438}},\ \bibinfo {pages}
  {197} (\bibinfo {year} {2005})}\BibitemShut {NoStop}%
\bibitem [{\citenamefont {Novoselov}\ \emph {et~al.}(2006)\citenamefont
  {Novoselov}, \citenamefont {McCann}, \citenamefont {Morozov}, \citenamefont
  {Fal'ko}, \citenamefont {Katsnelson}, \citenamefont {Zeitler}, \citenamefont
  {Jiang}, \citenamefont {Schedin},\ and\ \citenamefont
  {Geim}}]{NovoselovNP2006}%
  \BibitemOpen
  \bibfield  {author} {\bibinfo {author} {\bibfnamefont {K.~S.}\ \bibnamefont
  {Novoselov}}, \bibinfo {author} {\bibfnamefont {E.}~\bibnamefont {McCann}},
  \bibinfo {author} {\bibfnamefont {S.~V.}\ \bibnamefont {Morozov}}, \bibinfo
  {author} {\bibfnamefont {V.~I.}\ \bibnamefont {Fal'ko}}, \bibinfo {author}
  {\bibfnamefont {M.~I.}\ \bibnamefont {Katsnelson}}, \bibinfo {author}
  {\bibfnamefont {U.}~\bibnamefont {Zeitler}}, \bibinfo {author} {\bibfnamefont
  {D.}~\bibnamefont {Jiang}}, \bibinfo {author} {\bibfnamefont
  {F.}~\bibnamefont {Schedin}},\ and\ \bibinfo {author} {\bibfnamefont {A.~K.}\
  \bibnamefont {Geim}},\ }\href {https://doi.org/10.1038/nphys245} {\bibfield
  {journal} {\bibinfo  {journal} {Nature Physics}\ }\textbf {\bibinfo {volume}
  {2}},\ \bibinfo {pages} {177} (\bibinfo {year} {2006})}\BibitemShut {NoStop}%
\bibitem [{\citenamefont {Zhang}\ \emph {et~al.}(2005)\citenamefont {Zhang},
  \citenamefont {Tan}, \citenamefont {Stormer},\ and\ \citenamefont
  {Kim}}]{ZhangYBN2005}%
  \BibitemOpen
  \bibfield  {author} {\bibinfo {author} {\bibfnamefont {Y.-B.}\ \bibnamefont
  {Zhang}}, \bibinfo {author} {\bibfnamefont {Y.-W.}\ \bibnamefont {Tan}},
  \bibinfo {author} {\bibfnamefont {H.~L.}\ \bibnamefont {Stormer}},\ and\
  \bibinfo {author} {\bibfnamefont {P.}~\bibnamefont {Kim}},\ }\href
  {https://doi.org/10.1038/nature04235} {\bibfield  {journal} {\bibinfo
  {journal} {Nature}\ }\textbf {\bibinfo {volume} {438}},\ \bibinfo {pages}
  {201} (\bibinfo {year} {2005})}\BibitemShut {NoStop}%
\bibitem [{\citenamefont {Beenakker}(2008)}]{BeenakkerRMP2008}%
  \BibitemOpen
  \bibfield  {author} {\bibinfo {author} {\bibfnamefont {C.~W.~J.}\
  \bibnamefont {Beenakker}},\ }\href
  {https://doi.org/10.1103/RevModPhys.80.1337} {\bibfield  {journal} {\bibinfo
  {journal} {Rev. Mod. Phys.}\ }\textbf {\bibinfo {volume} {80}},\ \bibinfo
  {pages} {1337} (\bibinfo {year} {2008})}\BibitemShut {NoStop}%
\bibitem [{\citenamefont {Castro~Neto}\ \emph {et~al.}(2009)\citenamefont
  {Castro~Neto}, \citenamefont {Guinea}, \citenamefont {Peres}, \citenamefont
  {Novoselov},\ and\ \citenamefont {Geim}}]{NetoRMP2009}%
  \BibitemOpen
  \bibfield  {author} {\bibinfo {author} {\bibfnamefont {A.~H.}\ \bibnamefont
  {Castro~Neto}}, \bibinfo {author} {\bibfnamefont {F.}~\bibnamefont {Guinea}},
  \bibinfo {author} {\bibfnamefont {N.~M.~R.}\ \bibnamefont {Peres}}, \bibinfo
  {author} {\bibfnamefont {K.~S.}\ \bibnamefont {Novoselov}},\ and\ \bibinfo
  {author} {\bibfnamefont {A.~K.}\ \bibnamefont {Geim}},\ }\href
  {https://doi.org/10.1103/RevModPhys.81.109} {\bibfield  {journal} {\bibinfo
  {journal} {Rev. Mod. Phys.}\ }\textbf {\bibinfo {volume} {81}},\ \bibinfo
  {pages} {109} (\bibinfo {year} {2009})}\BibitemShut {NoStop}%
\bibitem [{\citenamefont {Brey}\ and\ \citenamefont
  {Fertig}(2006)}]{BLPRB2006}%
  \BibitemOpen
  \bibfield  {author} {\bibinfo {author} {\bibfnamefont {L.}~\bibnamefont
  {Brey}}\ and\ \bibinfo {author} {\bibfnamefont {H.~A.}\ \bibnamefont
  {Fertig}},\ }\href {https://doi.org/10.1103/PhysRevB.73.195408} {\bibfield
  {journal} {\bibinfo  {journal} {Phys. Rev. B}\ }\textbf {\bibinfo {volume}
  {73}},\ \bibinfo {pages} {195408} (\bibinfo {year} {2006})}\BibitemShut
  {NoStop}%
\bibitem [{\citenamefont {Abanin}\ \emph {et~al.}(2006)\citenamefont {Abanin},
  \citenamefont {Lee},\ and\ \citenamefont {Levitov}}]{ADAPRL2006}%
  \BibitemOpen
  \bibfield  {author} {\bibinfo {author} {\bibfnamefont {D.~A.}\ \bibnamefont
  {Abanin}}, \bibinfo {author} {\bibfnamefont {P.~A.}\ \bibnamefont {Lee}},\
  and\ \bibinfo {author} {\bibfnamefont {L.~S.}\ \bibnamefont {Levitov}},\
  }\href {https://doi.org/10.1103/PhysRevLett.96.176803} {\bibfield  {journal}
  {\bibinfo  {journal} {Phys. Rev. Lett.}\ }\textbf {\bibinfo {volume} {96}},\
  \bibinfo {pages} {176803} (\bibinfo {year} {2006})}\BibitemShut {NoStop}%
\bibitem [{\citenamefont {Fertig}\ and\ \citenamefont
  {Brey}(2006)}]{FHAPRL2006}%
  \BibitemOpen
  \bibfield  {author} {\bibinfo {author} {\bibfnamefont {H.~A.}\ \bibnamefont
  {Fertig}}\ and\ \bibinfo {author} {\bibfnamefont {L.}~\bibnamefont {Brey}},\
  }\href {https://doi.org/10.1103/PhysRevLett.97.116805} {\bibfield  {journal}
  {\bibinfo  {journal} {Phys. Rev. Lett.}\ }\textbf {\bibinfo {volume} {97}},\
  \bibinfo {pages} {116805} (\bibinfo {year} {2006})}\BibitemShut {NoStop}%
\bibitem [{\citenamefont {Kharitonov}\ \emph {et~al.}(2016)\citenamefont
  {Kharitonov}, \citenamefont {Juergens},\ and\ \citenamefont
  {Trauzettel}}]{KharitonovPRB2016}%
  \BibitemOpen
  \bibfield  {author} {\bibinfo {author} {\bibfnamefont {M.}~\bibnamefont
  {Kharitonov}}, \bibinfo {author} {\bibfnamefont {S.}~\bibnamefont
  {Juergens}},\ and\ \bibinfo {author} {\bibfnamefont {B.}~\bibnamefont
  {Trauzettel}},\ }\href {https://doi.org/10.1103/PhysRevB.94.035146}
  {\bibfield  {journal} {\bibinfo  {journal} {Phys. Rev. B}\ }\textbf {\bibinfo
  {volume} {94}},\ \bibinfo {pages} {035146} (\bibinfo {year}
  {2016})}\BibitemShut {NoStop}%
\bibitem [{\citenamefont {Veyrat}\ \emph {et~al.}(2020)\citenamefont {Veyrat},
  \citenamefont {Corentin}, \citenamefont {Coissard}, \citenamefont {Li},
  \citenamefont {Gay}, \citenamefont {Watanabe}, \citenamefont {Taniguchi},
  \citenamefont {Han}, \citenamefont {Piot}, \citenamefont {Sellier},\ and\
  \citenamefont {Sacépé}}]{VeyratScience2020}%
  \BibitemOpen
  \bibfield  {author} {\bibinfo {author} {\bibfnamefont {L.}~\bibnamefont
  {Veyrat}}, \bibinfo {author} {\bibfnamefont {D.}~\bibnamefont {Corentin}},
  \bibinfo {author} {\bibfnamefont {A.}~\bibnamefont {Coissard}}, \bibinfo
  {author} {\bibfnamefont {X.}~\bibnamefont {Li}}, \bibinfo {author}
  {\bibfnamefont {F.}~\bibnamefont {Gay}}, \bibinfo {author} {\bibfnamefont
  {K.}~\bibnamefont {Watanabe}}, \bibinfo {author} {\bibfnamefont
  {T.}~\bibnamefont {Taniguchi}}, \bibinfo {author} {\bibfnamefont
  {Z.}~\bibnamefont {Han}}, \bibinfo {author} {\bibfnamefont {B.}~\bibnamefont
  {Piot}}, \bibinfo {author} {\bibfnamefont {H.}~\bibnamefont {Sellier}},\ and\
  \bibinfo {author} {\bibfnamefont {B.}~\bibnamefont {Sacépé}},\ }\href
  {https://doi.org/10.1126/science.aax8201} {\bibfield  {journal} {\bibinfo
  {journal} {Science}\ }\textbf {\bibinfo {volume} {367}},\ \bibinfo {pages}
  {781} (\bibinfo {year} {2020})}\BibitemShut {NoStop}%
\bibitem [{\citenamefont {Sun}\ and\ \citenamefont {Xie}(2010)}]{QFSunPRL2010}%
  \BibitemOpen
  \bibfield  {author} {\bibinfo {author} {\bibfnamefont {Q.-F.}\ \bibnamefont
  {Sun}}\ and\ \bibinfo {author} {\bibfnamefont {X.~C.}\ \bibnamefont {Xie}},\
  }\href {https://doi.org/10.1103/PhysRevLett.104.066805} {\bibfield  {journal}
  {\bibinfo  {journal} {Phys. Rev. Lett.}\ }\textbf {\bibinfo {volume} {104}},\
  \bibinfo {pages} {066805} (\bibinfo {year} {2010})}\BibitemShut {NoStop}%
\bibitem [{\citenamefont {Gmitra}\ \emph {et~al.}(2009)\citenamefont {Gmitra},
  \citenamefont {Konschuh}, \citenamefont {Ertler}, \citenamefont
  {Ambrosch-Draxl},\ and\ \citenamefont {Fabian}}]{GMPRB2009}%
  \BibitemOpen
  \bibfield  {author} {\bibinfo {author} {\bibfnamefont {M.}~\bibnamefont
  {Gmitra}}, \bibinfo {author} {\bibfnamefont {S.}~\bibnamefont {Konschuh}},
  \bibinfo {author} {\bibfnamefont {C.}~\bibnamefont {Ertler}}, \bibinfo
  {author} {\bibfnamefont {C.}~\bibnamefont {Ambrosch-Draxl}},\ and\ \bibinfo
  {author} {\bibfnamefont {J.}~\bibnamefont {Fabian}},\ }\href
  {https://doi.org/10.1103/PhysRevB.80.235431} {\bibfield  {journal} {\bibinfo
  {journal} {Phys. Rev. B}\ }\textbf {\bibinfo {volume} {80}},\ \bibinfo
  {pages} {235431} (\bibinfo {year} {2009})}\BibitemShut {NoStop}%
\bibitem [{\citenamefont {Dedkov}\ \emph {et~al.}(2008)\citenamefont {Dedkov},
  \citenamefont {Fonin}, \citenamefont {R\"udiger},\ and\ \citenamefont
  {Laubschat}}]{DYSPRL2008}%
  \BibitemOpen
  \bibfield  {author} {\bibinfo {author} {\bibfnamefont {Y.~S.}\ \bibnamefont
  {Dedkov}}, \bibinfo {author} {\bibfnamefont {M.}~\bibnamefont {Fonin}},
  \bibinfo {author} {\bibfnamefont {U.}~\bibnamefont {R\"udiger}},\ and\
  \bibinfo {author} {\bibfnamefont {C.}~\bibnamefont {Laubschat}},\ }\href
  {https://doi.org/10.1103/PhysRevLett.100.107602} {\bibfield  {journal}
  {\bibinfo  {journal} {Phys. Rev. Lett.}\ }\textbf {\bibinfo {volume} {100}},\
  \bibinfo {pages} {107602} (\bibinfo {year} {2008})}\BibitemShut {NoStop}%
\bibitem [{\citenamefont {Jiang}\ \emph
  {et~al.}(2009{\natexlab{a}})\citenamefont {Jiang}, \citenamefont {Cheng},
  \citenamefont {Sun},\ and\ \citenamefont {Xie}}]{HuaJiangPRL2009}%
  \BibitemOpen
  \bibfield  {author} {\bibinfo {author} {\bibfnamefont {H.}~\bibnamefont
  {Jiang}}, \bibinfo {author} {\bibfnamefont {S.-G.}\ \bibnamefont {Cheng}},
  \bibinfo {author} {\bibfnamefont {Q.-F.}\ \bibnamefont {Sun}},\ and\ \bibinfo
  {author} {\bibfnamefont {X.~C.}\ \bibnamefont {Xie}},\ }\href
  {https://doi.org/10.1103/PhysRevLett.103.036803} {\bibfield  {journal}
  {\bibinfo  {journal} {Phys. Rev. Lett.}\ }\textbf {\bibinfo {volume} {103}},\
  \bibinfo {pages} {036803} (\bibinfo {year} {2009}{\natexlab{a}})}\BibitemShut
  {NoStop}%
\bibitem [{\citenamefont {Du}\ \emph {et~al.}(2015)\citenamefont {Du},
  \citenamefont {Knez}, \citenamefont {Sullivan},\ and\ \citenamefont
  {Du}}]{RRDuPRL2015}%
  \BibitemOpen
  \bibfield  {author} {\bibinfo {author} {\bibfnamefont {L.}~\bibnamefont
  {Du}}, \bibinfo {author} {\bibfnamefont {I.}~\bibnamefont {Knez}}, \bibinfo
  {author} {\bibfnamefont {G.}~\bibnamefont {Sullivan}},\ and\ \bibinfo
  {author} {\bibfnamefont {R.-R.}\ \bibnamefont {Du}},\ }\href
  {https://doi.org/10.1103/PhysRevLett.114.096802} {\bibfield  {journal}
  {\bibinfo  {journal} {Phys. Rev. Lett.}\ }\textbf {\bibinfo {volume} {114}},\
  \bibinfo {pages} {096802} (\bibinfo {year} {2015})}\BibitemShut {NoStop}%
\bibitem [{\citenamefont {Prodan}(2009)}]{PordanPRB2009}%
  \BibitemOpen
  \bibfield  {author} {\bibinfo {author} {\bibfnamefont {E.}~\bibnamefont
  {Prodan}},\ }\href {https://doi.org/10.1103/PhysRevB.80.125327} {\bibfield
  {journal} {\bibinfo  {journal} {Phys. Rev. B}\ }\textbf {\bibinfo {volume}
  {80}},\ \bibinfo {pages} {125327} (\bibinfo {year} {2009})}\BibitemShut
  {NoStop}%
\bibitem [{\citenamefont {Prodan}(2011)}]{Prodan2011}%
  \BibitemOpen
  \bibfield  {author} {\bibinfo {author} {\bibfnamefont {E.}~\bibnamefont
  {Prodan}},\ }\href {https://doi.org/10.1088/1751-8113/44/11/113001}
  {\bibfield  {journal} {\bibinfo  {journal} {Journal of Physics A:
  Mathematical and Theoretical}\ }\textbf {\bibinfo {volume} {44}},\ \bibinfo
  {pages} {113001} (\bibinfo {year} {2011})}\BibitemShut {NoStop}%
\bibitem [{\citenamefont {Datta}(1995)}]{datta1995}%
  \BibitemOpen
  \bibfield  {author} {\bibinfo {author} {\bibfnamefont {S.}~\bibnamefont
  {Datta}},\ }\href {https://doi.org/10.1017/CBO9780511805776} {\emph {\bibinfo
  {title} {Electronic Transport in Mesoscopic Systems}}},\ Cambridge Studies in
  Semiconductor Physics and Microelectronic Engineering\ (\bibinfo  {publisher}
  {Cambridge University Press},\ \bibinfo {year} {1995})\BibitemShut {NoStop}%
\bibitem [{\citenamefont {Huckestein}(1995)}]{RevModPhys.67.357}%
  \BibitemOpen
  \bibfield  {author} {\bibinfo {author} {\bibfnamefont {B.}~\bibnamefont
  {Huckestein}},\ }\href {https://doi.org/10.1103/RevModPhys.67.357} {\bibfield
   {journal} {\bibinfo  {journal} {Rev. Mod. Phys.}\ }\textbf {\bibinfo
  {volume} {67}},\ \bibinfo {pages} {357} (\bibinfo {year} {1995})}\BibitemShut
  {NoStop}%
\bibitem [{\citenamefont {Janssen}(1998)}]{JANSSEN19981}%
  \BibitemOpen
  \bibfield  {author} {\bibinfo {author} {\bibfnamefont {M.}~\bibnamefont
  {Janssen}},\ }\href
  {https://doi.org/https://doi.org/10.1016/S0370-1573(97)00050-1} {\bibfield
  {journal} {\bibinfo  {journal} {Physics Reports}\ }\textbf {\bibinfo {volume}
  {295}},\ \bibinfo {pages} {1} (\bibinfo {year} {1998})}\BibitemShut {NoStop}%
\bibitem [{\citenamefont {Pixley}\ \emph {et~al.}(2015)\citenamefont {Pixley},
  \citenamefont {Goswami},\ and\ \citenamefont {Das~Sarma}}]{PJH2015}%
  \BibitemOpen
  \bibfield  {author} {\bibinfo {author} {\bibfnamefont {J.~H.}\ \bibnamefont
  {Pixley}}, \bibinfo {author} {\bibfnamefont {P.}~\bibnamefont {Goswami}},\
  and\ \bibinfo {author} {\bibfnamefont {S.}~\bibnamefont {Das~Sarma}},\ }\href
  {https://doi.org/10.1103/PhysRevLett.115.076601} {\bibfield  {journal}
  {\bibinfo  {journal} {Phys. Rev. Lett.}\ }\textbf {\bibinfo {volume} {115}},\
  \bibinfo {pages} {076601} (\bibinfo {year} {2015})}\BibitemShut {NoStop}%
\bibitem [{\citenamefont {Wegner}(1980)}]{Wegner1980}%
  \BibitemOpen
  \bibfield  {author} {\bibinfo {author} {\bibfnamefont {F.}~\bibnamefont
  {Wegner}},\ }\href {https://doi.org/10.1007/BF01325284} {\bibfield  {journal}
  {\bibinfo  {journal} {Zeitschrift f{\"u}r Physik B Condensed Matter}\
  }\textbf {\bibinfo {volume} {36}},\ \bibinfo {pages} {209} (\bibinfo {year}
  {1980})}\BibitemShut {NoStop}%
\bibitem [{\citenamefont {Zhang}\ \emph
  {et~al.}(2009{\natexlab{a}})\citenamefont {Zhang}, \citenamefont {Hu},
  \citenamefont {Bernevig}, \citenamefont {Wang}, \citenamefont {Xie},\ and\
  \citenamefont {Liu}}]{ZYY2009}%
  \BibitemOpen
  \bibfield  {author} {\bibinfo {author} {\bibfnamefont {Y.-Y.}\ \bibnamefont
  {Zhang}}, \bibinfo {author} {\bibfnamefont {J.}~\bibnamefont {Hu}}, \bibinfo
  {author} {\bibfnamefont {B.~A.}\ \bibnamefont {Bernevig}}, \bibinfo {author}
  {\bibfnamefont {X.~R.}\ \bibnamefont {Wang}}, \bibinfo {author}
  {\bibfnamefont {X.~C.}\ \bibnamefont {Xie}},\ and\ \bibinfo {author}
  {\bibfnamefont {W.~M.}\ \bibnamefont {Liu}},\ }\href
  {https://doi.org/10.1103/PhysRevLett.102.106401} {\bibfield  {journal}
  {\bibinfo  {journal} {Phys. Rev. Lett.}\ }\textbf {\bibinfo {volume} {102}},\
  \bibinfo {pages} {106401} (\bibinfo {year} {2009}{\natexlab{a}})}\BibitemShut
  {NoStop}%
\bibitem [{\citenamefont {Edwards}\ and\ \citenamefont
  {Thouless}(1972)}]{JTEdwards1972}%
  \BibitemOpen
  \bibfield  {author} {\bibinfo {author} {\bibfnamefont {J.~T.}\ \bibnamefont
  {Edwards}}\ and\ \bibinfo {author} {\bibfnamefont {D.~J.}\ \bibnamefont
  {Thouless}},\ }\href {https://doi.org/10.1088/0022-3719/5/8/007} {\bibfield
  {journal} {\bibinfo  {journal} {Journal of Physics C: Solid State Physics}\
  }\textbf {\bibinfo {volume} {5}},\ \bibinfo {pages} {807} (\bibinfo {year}
  {1972})}\BibitemShut {NoStop}%
\bibitem [{\citenamefont {Li}\ \emph {et~al.}(2021)\citenamefont {Li},
  \citenamefont {Chen}, \citenamefont {Jiang},\ and\ \citenamefont
  {Xie}}]{LHL2021}%
  \BibitemOpen
  \bibfield  {author} {\bibinfo {author} {\bibfnamefont {H.}~\bibnamefont
  {Li}}, \bibinfo {author} {\bibfnamefont {C.-Z.}\ \bibnamefont {Chen}},
  \bibinfo {author} {\bibfnamefont {H.}~\bibnamefont {Jiang}},\ and\ \bibinfo
  {author} {\bibfnamefont {X.~C.}\ \bibnamefont {Xie}},\ }\href
  {https://doi.org/10.1103/PhysRevLett.127.236402} {\bibfield  {journal}
  {\bibinfo  {journal} {Phys. Rev. Lett.}\ }\textbf {\bibinfo {volume} {127}},\
  \bibinfo {pages} {236402} (\bibinfo {year} {2021})}\BibitemShut {NoStop}%
\bibitem [{\citenamefont {Jiang}\ \emph
  {et~al.}(2009{\natexlab{b}})\citenamefont {Jiang}, \citenamefont {Wang},
  \citenamefont {Sun},\ and\ \citenamefont {Xie}}]{HuaJiangPRB2009}%
  \BibitemOpen
  \bibfield  {author} {\bibinfo {author} {\bibfnamefont {H.}~\bibnamefont
  {Jiang}}, \bibinfo {author} {\bibfnamefont {L.}~\bibnamefont {Wang}},
  \bibinfo {author} {\bibfnamefont {Q.-F.}\ \bibnamefont {Sun}},\ and\ \bibinfo
  {author} {\bibfnamefont {X.~C.}\ \bibnamefont {Xie}},\ }\href
  {https://doi.org/10.1103/PhysRevB.80.165316} {\bibfield  {journal} {\bibinfo
  {journal} {Phys. Rev. B}\ }\textbf {\bibinfo {volume} {80}},\ \bibinfo
  {pages} {165316} (\bibinfo {year} {2009}{\natexlab{b}})}\BibitemShut
  {NoStop}%
\bibitem [{\citenamefont {Zhang}\ \emph {et~al.}(2019)\citenamefont {Zhang},
  \citenamefont {Wu}, \citenamefont {Song},\ and\ \citenamefont
  {Jiang}}]{ZZQPRB2019}%
  \BibitemOpen
  \bibfield  {author} {\bibinfo {author} {\bibfnamefont {Z.-Q.}\ \bibnamefont
  {Zhang}}, \bibinfo {author} {\bibfnamefont {B.-L.}\ \bibnamefont {Wu}},
  \bibinfo {author} {\bibfnamefont {J.}~\bibnamefont {Song}},\ and\ \bibinfo
  {author} {\bibfnamefont {H.}~\bibnamefont {Jiang}},\ }\href
  {https://doi.org/10.1103/PhysRevB.100.184202} {\bibfield  {journal} {\bibinfo
   {journal} {Phys. Rev. B}\ }\textbf {\bibinfo {volume} {100}},\ \bibinfo
  {pages} {184202} (\bibinfo {year} {2019})}\BibitemShut {NoStop}%
\bibitem [{\citenamefont {Zhang}\ \emph {et~al.}(2021)\citenamefont {Zhang},
  \citenamefont {Chen}, \citenamefont {Wu}, \citenamefont {Jiang},
  \citenamefont {Liu}, \citenamefont {Sun},\ and\ \citenamefont
  {Xie}}]{ZZQPRB2021}%
  \BibitemOpen
  \bibfield  {author} {\bibinfo {author} {\bibfnamefont {Z.-Q.}\ \bibnamefont
  {Zhang}}, \bibinfo {author} {\bibfnamefont {C.-Z.}\ \bibnamefont {Chen}},
  \bibinfo {author} {\bibfnamefont {Y.}~\bibnamefont {Wu}}, \bibinfo {author}
  {\bibfnamefont {H.}~\bibnamefont {Jiang}}, \bibinfo {author} {\bibfnamefont
  {J.}~\bibnamefont {Liu}}, \bibinfo {author} {\bibfnamefont {Q.-f.}\
  \bibnamefont {Sun}},\ and\ \bibinfo {author} {\bibfnamefont {X.~C.}\
  \bibnamefont {Xie}},\ }\href {https://doi.org/10.1103/PhysRevB.103.075434}
  {\bibfield  {journal} {\bibinfo  {journal} {Phys. Rev. B}\ }\textbf {\bibinfo
  {volume} {103}},\ \bibinfo {pages} {075434} (\bibinfo {year}
  {2021})}\BibitemShut {NoStop}%
\bibitem [{\citenamefont {MacKinnon}\ and\ \citenamefont
  {Kramer}(1981)}]{MacKinnon1981}%
  \BibitemOpen
  \bibfield  {author} {\bibinfo {author} {\bibfnamefont {A.}~\bibnamefont
  {MacKinnon}}\ and\ \bibinfo {author} {\bibfnamefont {B.}~\bibnamefont
  {Kramer}},\ }\href {https://doi.org/10.1103/PhysRevLett.47.1546} {\bibfield
  {journal} {\bibinfo  {journal} {Phys. Rev. Lett.}\ }\textbf {\bibinfo
  {volume} {47}},\ \bibinfo {pages} {1546} (\bibinfo {year}
  {1981})}\BibitemShut {NoStop}%
\bibitem [{\citenamefont {MacKinnon}\ and\ \citenamefont
  {Kramer}(1983)}]{MacKinnon1983}%
  \BibitemOpen
  \bibfield  {author} {\bibinfo {author} {\bibfnamefont {A.}~\bibnamefont
  {MacKinnon}}\ and\ \bibinfo {author} {\bibfnamefont {B.}~\bibnamefont
  {Kramer}},\ }\href {https://doi.org/10.1007/BF01578242} {\bibfield  {journal}
  {\bibinfo  {journal} {Zeitschrift für Physik B Condensed Matter}\ }\textbf
  {\bibinfo {volume} {53}},\ \bibinfo {pages} {1} (\bibinfo {year}
  {1983})}\BibitemShut {NoStop}%
\bibitem [{\citenamefont {Kramer}\ and\ \citenamefont
  {MacKinnon}(1993)}]{Kramer1993}%
  \BibitemOpen
  \bibfield  {author} {\bibinfo {author} {\bibfnamefont {B.}~\bibnamefont
  {Kramer}}\ and\ \bibinfo {author} {\bibfnamefont {A.}~\bibnamefont
  {MacKinnon}},\ }\href {https://doi.org/10.1088/0034-4885/56/12/001}
  {\bibfield  {journal} {\bibinfo  {journal} {Reports on Progress in Physics}\
  }\textbf {\bibinfo {volume} {56}},\ \bibinfo {pages} {1469} (\bibinfo {year}
  {1993})}\BibitemShut {NoStop}%
\bibitem [{\citenamefont {Beenakker}(1997)}]{Beenakker1997}%
  \BibitemOpen
  \bibfield  {author} {\bibinfo {author} {\bibfnamefont {C.~W.~J.}\
  \bibnamefont {Beenakker}},\ }\href
  {https://doi.org/10.1103/RevModPhys.69.731} {\bibfield  {journal} {\bibinfo
  {journal} {Rev. Mod. Phys.}\ }\textbf {\bibinfo {volume} {69}},\ \bibinfo
  {pages} {731} (\bibinfo {year} {1997})}\BibitemShut {NoStop}%
\bibitem [{\citenamefont {Xiong}\ and\ \citenamefont
  {Xiong}(2007)}]{SJXiong2007}%
  \BibitemOpen
  \bibfield  {author} {\bibinfo {author} {\bibfnamefont {S.-J.}\ \bibnamefont
  {Xiong}}\ and\ \bibinfo {author} {\bibfnamefont {Y.}~\bibnamefont {Xiong}},\
  }\href {https://doi.org/10.1103/PhysRevB.76.214204} {\bibfield  {journal}
  {\bibinfo  {journal} {Phys. Rev. B}\ }\textbf {\bibinfo {volume} {76}},\
  \bibinfo {pages} {214204} (\bibinfo {year} {2007})}\BibitemShut {NoStop}%
\bibitem [{\citenamefont {Zhang}\ \emph
  {et~al.}(2009{\natexlab{b}})\citenamefont {Zhang}, \citenamefont {Hu},
  \citenamefont {Bernevig}, \citenamefont {Wang}, \citenamefont {Xie},\ and\
  \citenamefont {Liu}}]{YYZhangPRL2009}%
  \BibitemOpen
  \bibfield  {author} {\bibinfo {author} {\bibfnamefont {Y.-Y.}\ \bibnamefont
  {Zhang}}, \bibinfo {author} {\bibfnamefont {J.}~\bibnamefont {Hu}}, \bibinfo
  {author} {\bibfnamefont {B.~A.}\ \bibnamefont {Bernevig}}, \bibinfo {author}
  {\bibfnamefont {X.~R.}\ \bibnamefont {Wang}}, \bibinfo {author}
  {\bibfnamefont {X.~C.}\ \bibnamefont {Xie}},\ and\ \bibinfo {author}
  {\bibfnamefont {W.~M.}\ \bibnamefont {Liu}},\ }\href
  {https://doi.org/10.1103/PhysRevLett.102.106401} {\bibfield  {journal}
  {\bibinfo  {journal} {Phys. Rev. Lett.}\ }\textbf {\bibinfo {volume} {102}},\
  \bibinfo {pages} {106401} (\bibinfo {year} {2009}{\natexlab{b}})}\BibitemShut
  {NoStop}%
\bibitem [{\citenamefont {Sun}\ and\ \citenamefont
  {Xie}(2009)}]{QFSunJPCM2009}%
  \BibitemOpen
  \bibfield  {author} {\bibinfo {author} {\bibfnamefont {Q.-F.}\ \bibnamefont
  {Sun}}\ and\ \bibinfo {author} {\bibfnamefont {X.}~\bibnamefont {Xie}},\
  }\href {https://doi.org/10.1088/0953-8984/21/34/344204} {\bibfield  {journal}
  {\bibinfo  {journal} {Journal of physics. Condensed matter : an Institute of
  Physics journal}\ }\textbf {\bibinfo {volume} {21}},\ \bibinfo {pages}
  {344204} (\bibinfo {year} {2009})}\BibitemShut {NoStop}%
\bibitem [{\citenamefont {Chen}\ \emph
  {et~al.}(2018{\natexlab{a}})\citenamefont {Chen}, \citenamefont {He},
  \citenamefont {Ali}, \citenamefont {Lee}, \citenamefont {Fong},\ and\
  \citenamefont {Law}}]{CZChenPRB2018}%
  \BibitemOpen
  \bibfield  {author} {\bibinfo {author} {\bibfnamefont {C.-Z.}\ \bibnamefont
  {Chen}}, \bibinfo {author} {\bibfnamefont {J.~J.}\ \bibnamefont {He}},
  \bibinfo {author} {\bibfnamefont {M.~N.}\ \bibnamefont {Ali}}, \bibinfo
  {author} {\bibfnamefont {G.-H.}\ \bibnamefont {Lee}}, \bibinfo {author}
  {\bibfnamefont {K.~C.}\ \bibnamefont {Fong}},\ and\ \bibinfo {author}
  {\bibfnamefont {K.~T.}\ \bibnamefont {Law}},\ }\href
  {https://doi.org/10.1103/PhysRevB.98.075430} {\bibfield  {journal} {\bibinfo
  {journal} {Phys. Rev. B}\ }\textbf {\bibinfo {volume} {98}},\ \bibinfo
  {pages} {075430} (\bibinfo {year} {2018}{\natexlab{a}})}\BibitemShut
  {NoStop}%
\bibitem [{\citenamefont {Chen}\ \emph
  {et~al.}(2018{\natexlab{b}})\citenamefont {Chen}, \citenamefont {He},
  \citenamefont {Xu},\ and\ \citenamefont {Law}}]{CZChenPRB20182}%
  \BibitemOpen
  \bibfield  {author} {\bibinfo {author} {\bibfnamefont {C.-Z.}\ \bibnamefont
  {Chen}}, \bibinfo {author} {\bibfnamefont {J.~J.}\ \bibnamefont {He}},
  \bibinfo {author} {\bibfnamefont {D.-H.}\ \bibnamefont {Xu}},\ and\ \bibinfo
  {author} {\bibfnamefont {K.~T.}\ \bibnamefont {Law}},\ }\href
  {https://doi.org/10.1103/PhysRevB.98.165439} {\bibfield  {journal} {\bibinfo
  {journal} {Phys. Rev. B}\ }\textbf {\bibinfo {volume} {98}},\ \bibinfo
  {pages} {165439} (\bibinfo {year} {2018}{\natexlab{b}})}\BibitemShut
  {NoStop}%
\bibitem [{\citenamefont {Liu}\ \emph {et~al.}(2014)\citenamefont {Liu},
  \citenamefont {Wang},\ and\ \citenamefont {Zhang}}]{JieLiuPRB2014}%
  \BibitemOpen
  \bibfield  {author} {\bibinfo {author} {\bibfnamefont {J.}~\bibnamefont
  {Liu}}, \bibinfo {author} {\bibfnamefont {J.}~\bibnamefont {Wang}},\ and\
  \bibinfo {author} {\bibfnamefont {F.-C.}\ \bibnamefont {Zhang}},\ }\href
  {https://doi.org/10.1103/PhysRevB.90.035307} {\bibfield  {journal} {\bibinfo
  {journal} {Phys. Rev. B}\ }\textbf {\bibinfo {volume} {90}},\ \bibinfo
  {pages} {035307} (\bibinfo {year} {2014})}\BibitemShut {NoStop}%
\bibitem [{\citenamefont {Liu}\ \emph {et~al.}(2017)\citenamefont {Liu},
  \citenamefont {Liu}, \citenamefont {Song}, \citenamefont {Sun},\ and\
  \citenamefont {Xie}}]{JieLiuPRB2017}%
  \BibitemOpen
  \bibfield  {author} {\bibinfo {author} {\bibfnamefont {J.}~\bibnamefont
  {Liu}}, \bibinfo {author} {\bibfnamefont {H.}~\bibnamefont {Liu}}, \bibinfo
  {author} {\bibfnamefont {J.}~\bibnamefont {Song}}, \bibinfo {author}
  {\bibfnamefont {Q.-F.}\ \bibnamefont {Sun}},\ and\ \bibinfo {author}
  {\bibfnamefont {X.~C.}\ \bibnamefont {Xie}},\ }\href
  {https://doi.org/10.1103/PhysRevB.96.045401} {\bibfield  {journal} {\bibinfo
  {journal} {Phys. Rev. B}\ }\textbf {\bibinfo {volume} {96}},\ \bibinfo
  {pages} {045401} (\bibinfo {year} {2017})}\BibitemShut {NoStop}%
\bibitem [{\citenamefont {Law}\ \emph {et~al.}(2009)\citenamefont {Law},
  \citenamefont {Lee},\ and\ \citenamefont {Ng}}]{KTLawPRL2009}%
  \BibitemOpen
  \bibfield  {author} {\bibinfo {author} {\bibfnamefont {K.~T.}\ \bibnamefont
  {Law}}, \bibinfo {author} {\bibfnamefont {P.~A.}\ \bibnamefont {Lee}},\ and\
  \bibinfo {author} {\bibfnamefont {T.~K.}\ \bibnamefont {Ng}},\ }\href
  {https://doi.org/10.1103/PhysRevLett.103.237001} {\bibfield  {journal}
  {\bibinfo  {journal} {Phys. Rev. Lett.}\ }\textbf {\bibinfo {volume} {103}},\
  \bibinfo {pages} {237001} (\bibinfo {year} {2009})}\BibitemShut {NoStop}%
\bibitem [{\citenamefont {Fidkowski}\ \emph {et~al.}(2012)\citenamefont
  {Fidkowski}, \citenamefont {Alicea}, \citenamefont {Lindner}, \citenamefont
  {Lutchyn},\ and\ \citenamefont {Fisher}}]{FidkowskiPRB2012}%
  \BibitemOpen
  \bibfield  {author} {\bibinfo {author} {\bibfnamefont {L.}~\bibnamefont
  {Fidkowski}}, \bibinfo {author} {\bibfnamefont {J.}~\bibnamefont {Alicea}},
  \bibinfo {author} {\bibfnamefont {N.~H.}\ \bibnamefont {Lindner}}, \bibinfo
  {author} {\bibfnamefont {R.~M.}\ \bibnamefont {Lutchyn}},\ and\ \bibinfo
  {author} {\bibfnamefont {M.~P.~A.}\ \bibnamefont {Fisher}},\ }\href
  {https://doi.org/10.1103/PhysRevB.85.245121} {\bibfield  {journal} {\bibinfo
  {journal} {Phys. Rev. B}\ }\textbf {\bibinfo {volume} {85}},\ \bibinfo
  {pages} {245121} (\bibinfo {year} {2012})}\BibitemShut {NoStop}%
\bibitem [{\citenamefont {Haim}\ \emph {et~al.}(2019)\citenamefont {Haim},
  \citenamefont {Ilan},\ and\ \citenamefont {Alicea}}]{HaimPRL2019}%
  \BibitemOpen
  \bibfield  {author} {\bibinfo {author} {\bibfnamefont {A.}~\bibnamefont
  {Haim}}, \bibinfo {author} {\bibfnamefont {R.}~\bibnamefont {Ilan}},\ and\
  \bibinfo {author} {\bibfnamefont {J.}~\bibnamefont {Alicea}},\ }\href
  {https://doi.org/10.1103/PhysRevLett.123.046801} {\bibfield  {journal}
  {\bibinfo  {journal} {Phys. Rev. Lett.}\ }\textbf {\bibinfo {volume} {123}},\
  \bibinfo {pages} {046801} (\bibinfo {year} {2019})}\BibitemShut {NoStop}%
\end{thebibliography}%

\end{document}